%% file: main.tex
\documentclass{jfm}
\input{header.tex}
\usepackage{graphicx}
\usepackage{epstopdf, epsfig}

\shorttitle{DRL for sensor placement and robust control}
\shortauthor{R. Paris, S. Beneddine and J. Dandois}

\title{Robust flow control and optimal sensor placement using deep reinforcement learning}

\author{Romain Paris\aff{1}
  \corresp{\email{romain.paris@onera.fr}},
  Samir Beneddine\aff{1}
 \and Julien Dandois\aff{1}}

\affiliation{\aff{1}ONERA DAAA, 8 rue des Vertugadins, 92190 Meudon, France}

\begin{document}

\maketitle

\input{P0_Intro}
\input{P1_Methods}
\input{P2_SPPOCMA}
\input{P3_Results}
\input{P4_Conclusion}
\input{Z_Appendices}
\bibliographystyle{jfm}
\bibliography{biblioThese}

\end{document}

%% file: header.tex
\usepackage{amsmath}
\usepackage{amsfonts}
\usepackage{bbm}

\DeclareMathOperator*{\argmin}{arg\,min}
\newcommand{\Expec}[2]{\mathbb{E}_{#1}\left[#2\right]}
\let\oldforall\forall
\renewcommand{\forall}{\;\oldforall\,}
\newcommand{\pd}[2]{\frac{\partial #1}{\partial #2}}

\newcommand{\ident}{\mathcal{I}}

\renewcommand{\vec}[1]{\boldsymbol{#1}}
\newcommand{\mat}[1]{\mathsfbi{#1}}

\usepackage{listings}

\newcommand\pythonstyle{\lstset{
language=Python,
basicstyle=\ttm,
otherkeywords={},             
keywordstyle=\ttb\color{deepblue},
commentstyle=\ttm\color{deepgreen},
emph={rl,rlFramework},          
emphstyle=\ttb\color{deepred},    
stringstyle=\color{deepgreen},
frame=tb,                         
showstringspaces=false            %
}}

\lstnewenvironment{python}[1][]
{
\pythonstyle
\lstset{#1}
}
{}

\newcommand\pythoninline[1]{{\pythonstyle\lstinline!#1!}}

\usepackage{multirow}

\usepackage[makeroom]{cancel}

\usepackage{algorithm}
\usepackage{algorithmic}

\usepackage{pdfpages}


%% file: P0_Intro.tex
\begin{abstract}
This paper focuses on a drag-reducing control strategy on a 2D-simulated laminar flow past a cylinder. Deep reinforcement learning algorithms have been implemented to discover efficient control schemes, using two synthetic jets located on the cylinder's poles as actuators and pressure sensors in the wake of the cylinder as feedback observation. The present work focuses on the efficiency and robustness of the identified control strategy and introduces a novel algorithm (S-PPO-CMA) to optimise the sensor layout. An energy-efficient control strategy reducing drag by $18.4\%$ at Reynolds number $120$ is obtained. This control policy is shown to be robust both to the Reynolds number in the range $[100,216]$ and to measurement noise, enduring signal to noise ratios as low as $0.2$ with negligible impact on performance. Along with a systematic study on sensor number and location, the proposed sparsity-seeking algorithm has achieved a successful optimisation to a reduced 5-sensor layout while keeping state-of-the-art performance. These results highlight the interesting possibilities of reinforcement learning for active flow control and pave the way to efficient, robust and practical implementations of these control techniques in experimental or industrial systems.
\end{abstract} 

\section{Introduction}
Improvement of aerodynamic characteristics on air vehicles has mainly been achieved through shape optimisation in the past decades, with drag reduction as the primary goal. Passive control devices have long been the centrepiece of flow control \citep{Selby1992,Gutmark1999,Marquet2008}, thanks to their ease of use. Yet, their overall low efficiency advocates for active forms of control, which split into two categories: open-loop strategies (see for instance \citet{Sipp2012}), and closed-loop approaches \citep{Sipp2016}, the latter being known to display greater performance and robustness, taking advantage of state measurements. In this context, linear techniques have recently been investigated for active flow control. But they have shown some limitations on nonlinear systems, whether on performance, robustness, or computation complexity \citep{SippMarquetMeligaEtAl2010}.

These linear approaches often rely on a reduced-order model, mainly via proper orthogonal decomposition (POD) \citep{Gerhard2003,Bergmann2005} or resolvent analyses \citep{LeclercqDemourantPoussot-VassalEtAl2019}, which provide frameworks to apply linear control techniques. Numerous studies and various approaches have been proposed: \citet{Fujisawa2001}  and \citet{Siegel2003} used variable phase proportional and differential control to reduce the drag of low Reynolds number cylinder flows. Several studies implemented robust $H_2$ or $H_\infty$ control methods based on resolvent analysis \citep{Jin2019} or using an iterative strategy \citep{LeclercqDemourantPoussot-VassalEtAl2019}. Other mathematical frameworks, such as the adjoint approach \citep{He2000}, have also shown efficient control but at a high computational cost. 
These are a few examples of a large body of work dedicated to techniques that rely on local linear approximations, and thus, often pertain to constant or periodic forcing on the flow. Such control strategies are adapted to weakly nonlinear systems where the linear approach remains valid. They have been nonetheless often applied on nonlinear systems, despite their limitations, due to the lack of robust and efficient methods to tackle nonlinear high-dimensional systems, such as encountered in actual fluid mechanics applications.

In the context of the development of new and promising machine learning (ML) techniques, efficient nonlinear active flow control appears increasingly viable, as emphasised by \citet{BruntonNoack2015} and \citet{Brunton2020}. The use of artificial neural networks (NN) as universal function approximators that can be trained efficiently has already proven significant capabilities for solving complex problems such as translation \citep{Cho2014,Sutskever2014} or image recognition \citep{He2016}. Coupled with reinforcement methods, that use interactions with the controlled environment to improve performance, these techniques achieve autonomous learning of complex tasks \citep{Baker2019,Kaiser2019}, and often perform better than human experts \citep{MnihKavukcuogluSilverEtAl2015}. The present study focuses particularly on \textit{on-policy} Deep Reinforcement Learning (DRL). From the first methods \citep{Williams1992} to the most recent algorithms such as Trust Region Policy Optimisation \citep{SchulmanLevineAbbeelEtAl2015} or Proximal Policy Optimisation (PPO) and its variants \citep{SchulmanWolskiDhariwalEtAl2017,HaemaelaeinenBabadiMaEtAl2018}, DRL has demonstrated the ability to efficiently learn non-trivial control strategies (named policies) in complex and high-dimensional environments. By leveraging stochastic estimation and Markov processes, these algorithms optimise both sample and policy efficiencies. 

The use of ML techniques in fluid mechanics enables efficient and more straightforward nonlinear control strategies. In the wake of nonlinear auto-regressive models \citep{Kim2006,DandoisGarnierPamart2013}, ML algorithms are used either for black-box or model-based feedback control \citep{Seidel2009,Cohen2012}, leveraging the flexibility of neural network structures, using for instance the artificial neural network estimator (ANNE) method \citep{Noergaard2000}. Semi-supervised learning methods, such as model predictive control \citep{Nair2020} or DRL for control \citep{RabaultKuchtaJensenEtAl2019}, also meet increasing success. However, neural network's known lack of extrapolation capabilities and thus weak robustness for supervised learning applications highlights the robustness of DRL methods as a potential weak point. This issue is explored by \citet{Tang2020} who successfully controlled a 2D-cylinder wake across a large range of Reynolds numbers. In this study, the potentialities of deep reinforcement learning in active flow control performance, power efficiency and robustness are investigated on a similar test case. One important contribution of the paper relates to the introduction of a variant of PPO which is shown to outperform state-ot-the-art DRL approaches on the cylinder case.

Another important contribution relates to optimal sensor placement. Reducing sensor requirements while keeping optimal control performances is key to the potential transposition of these techniques to experimental and industrial cases. \citet{RabaultKuchtaJensenEtAl2019} and \citet{Tang2020} respectively use 151 and 236 probes to control the flow past a 2D-cylinder. Their work is therefore a first step for DRL control that needs to be continued and further improved, which is precisely the purpose of this work. The present study builds on these existing papers to propose new techniques and algorithms that reduce the gap between DRL capabilities for flow control and the requirements for a future experimental or industrial implementation. Having less sensors means at the same time less hardware requirements, less potential failure modes and less computational power needs, especially in a context of embedded systems with strong real-time computing constraints. This issue of optimal measurement location has been investigated by many authors outside of the context of DRL, for instance by \citet{Mons2017,Mons2016,Foures2014,Verma2020} for data assimilation. \citet{Bright2013} took advantage of compressed sensing to perform flow reconstruction using a limited number of sensors. The optimal estimation of a reduced order state, usually POD modes, has been used by \citet{Cohen2006,Seidel2009} and echoes the assumption that accurate flow estimation is an essential feature of efficient control. However, as stressed by \citet{Oehler2018}, control does not systematically require faithful flow reconstruction (in the sense of POD), the partial knowledge of relevant "hidden" variables may be sufficient. This idea is, in the linear framework, conveyed by the notion of observability Gramian. Empirical observability Gramians were used by \citet{Singh2005,DeVries2013} for flow estimation and a version balancing both observability and controlability Gramians was successfully implemented by \citet{Manohar2018} to design an optimal $H_2$ control of a linearised Ginzburg-Landau model. One of the main contributions of our work is to propose a new method, leveraging a previously learned DRL policy to optimise sensor location.

In the following, the simulated case study is first introduced, then the DRL algorithm used to derive the control strategy is described and a new sparsity-seeking variant of this algorithm, aiming at optimising sensor number and location is proposed. The main results and discussions are developed in section \ref{sec:results}. The issue of control efficiency and robustness to both Reynolds number variations and measurement noise is addressed in this section. Finally, the optimal sensor layouts derived using our proposed method is discussed.

%% file: P1_Methods.tex
\section{Description of the flow configuration and numerical methods}
\begin{figure}
    \centerline{
    \includegraphics[page=9,width=.35\textwidth]{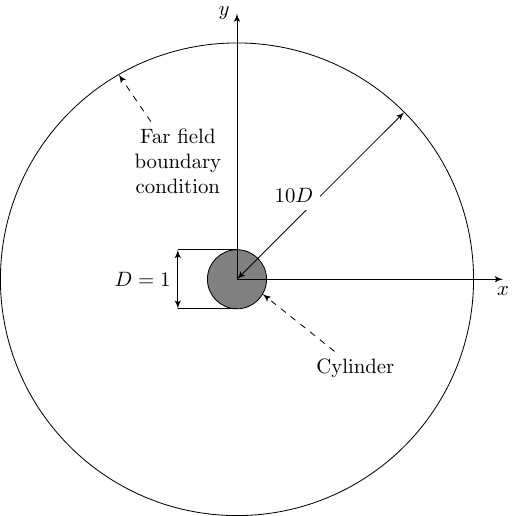}
    \includegraphics[page=10,width=.35\textwidth]{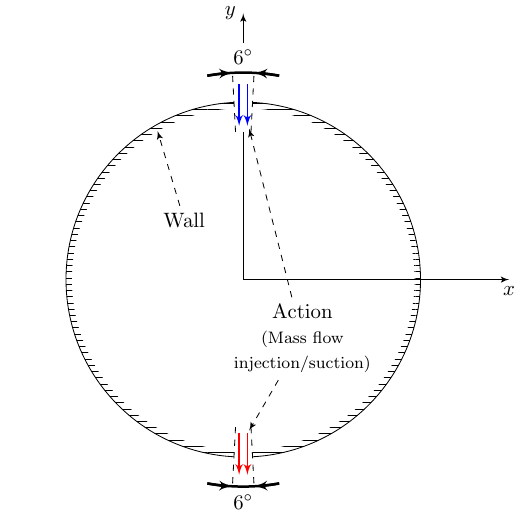}
    }
    \caption{Flow domain geometry. (left): Full domain, not at true scale. The far field boundary condition is a characteristic-based inflow/outflow boundary condition modelling free-stream flow. (right): Boundary conditions on the cylinder.}
    \label{fig:setup}
\end{figure}

The studied configuration is a bi-dimensional (2D) flow past a cylinder. The geometry is made non-dimensional by setting the cylinder diameter $D$ to 1. The centre of the cylinder is located at the origin $(0,0)$ of the flow domain. Figure \ref{fig:setup} displays the computed flow domain, which spans over $10D$, and shows the orientation of axes $x$ and $y$. 

\subsection{Numerical setup}
\label{sec:numericalSetup}
The flow is described by the compressible Navier-Stokes equations. The free-stream flow is uniform at a Mach number $M_\infty$ of $0.15$, oriented along $x$. In the following, all quantities are made non-dimensional by the characteristic length $D$, the inflow density $\rho_\infty$, the velocity $U_\infty$ and the static temperature $T_\infty$. Note that the Mach number is very low, such that the density fluctuations in the whole domain are negligible, and the flow is therefore quasi-incompressible. The Reynolds number $\Rey$, defined as $U_\infty D/\nu$ ($\nu$ being the kinematic viscosity), is varied in the article, but the first sections of the paper focus on a reference configuration at $\Rey=120$. The flow field is computed using ONERA's FastS finite volume method solver \citep{Dandois2018} for both steady and unsteady computations. For unsteady computations, a global numerical time step $dt~=~5~\times~10^{-3}$ is chosen. Additional numerical details are available in Appendix \ref{app:params}. The C-shaped structured mesh is made of 25200 nodes and is refined in the vicinity of the cylinder. The boundary conditions are specified in figure \ref{fig:setup}.

In the context of active flow control, and as shown in figure \ref{fig:setup}, injection or suction is performed on the cylinder's poles through two $6^\circ$-wide jet inlets. A control step $\Delta t$ is defined as the number of numerical iterations during which the control command is held constant. In this study a control step lasts for $50$ numerical time steps, thus $\Delta t=0.25$ non-dimensional time units. For each control step, an action command $a_t$ (positive or negative) is translated into a blowing/suction using a 20-iteration interpolation ramp in order to avoid abrupt changes of boundary conditions and non-physical values due to the numerical schemes, similarly to what \citet{RabaultKuchtaJensenEtAl2019} did. Formally, for the $i^{th}$ numerical iteration of the control step, the mass flow per unit area $q_i$ is:
\begin{equation}
  q_i = \rho_\infty U_\infty\left(a_{t-1}(1-r_i) + a_tr_i\right), \mbox{ with }   r_i = \left\{\begin{matrix}i/20& \text{ if } i<20\\1&\text{ otherwise}\end{matrix}\right.
\end{equation}

To ensure an instantaneous zero-net-mass-flux for every action, the two poles act reciprocally: $+q_i$ is imposed on the top inlet surface and $-q_i$ on the bottom inlet. Note that in several studies, the actuators are such that they are able to inject streamwise momentum, which may directly reduce the cylinder drag. In the present study, the actuators are designed such that they can only inject cross-stream momentum, thus making any "direct" drag reduction impossible.

\begin{figure}
    \centerline{
    \includegraphics[trim={2cm 0 0 0},clip,width=\textwidth]{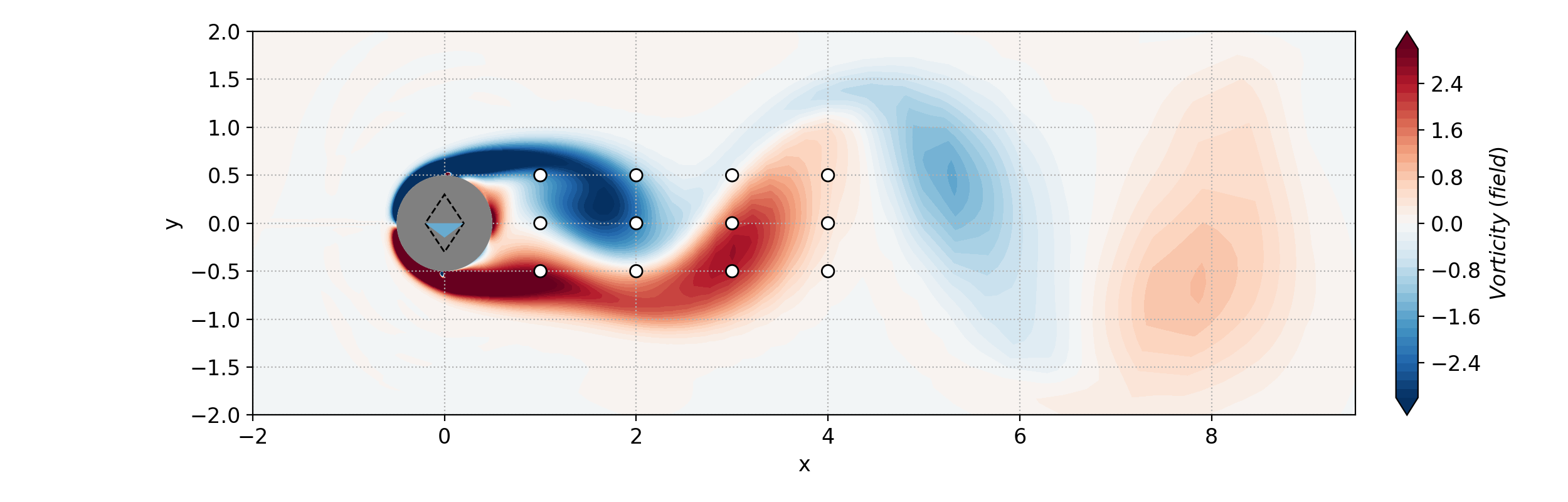}
    }
    \caption{Instantaneous vorticity flow field with action $a_t=-0.15$ at $\Rey=120$. White dots represent the sensor locations. The coloured triangle in the cylinder depicts the action, its height and colour representing its amplitude. The dashed diamond shape marks off maximum actions (both positive and negative).}
    \label{fig:refCase}
\end{figure}

Several sensors record the pressure of the flow at predefined locations at the end of every control step. The output measurement is a pressure fluctuation, defined as the difference between the local non-dimensional static pressure and the reference inflow static pressure $p_\infty$. Figure \ref{fig:refCase} illustrates a standard setup for this case.

Both drag and lift coefficients ($C_x$ and $C_l$) are computed on the cylinder via the resulting force of the flow $\vec{F}$:
\begin{equation}
    \vec{F} = \int_{\text{cylinder}}\mat{\sigma}.{\vec{n}}dS
\end{equation}
\begin{equation}
    C_x = \frac{\vec{F}.\vec{e}_x}{\frac{1}{2}\rho_\infty U_\infty^2 D}
\end{equation}
\begin{equation}
    C_l = \frac{\vec{F}.\vec{e}_y}{\frac{1}{2}\rho_\infty U_\infty^2 D}
\end{equation}
    where $\vec{n}$ is the unitary cylinder surface normal vector, $\mat{\sigma}$ is the stress tensor, $\vec{e}_x = (1,0)$ and $\vec{e}_y = (0,1)$.


\subsection{Uncontrolled flow}\label{sec:unc_flow}
The uncontrolled configuration, denoted in the following as the baseline flow, displays a well-documented vortex shedding behaviour \citep{Williamson1996} that appears for Reynolds numbers above 46, and which is due to a Hopf bifurcation where the steady solution of the Navier-Stokes equations (the base flow) becomes unstable. Thus, the flow becomes unsteady and follows a stable limit cycle associated with vortex shedding. 

\begin{figure}
    \centerline{
    \includegraphics[width=\textwidth]{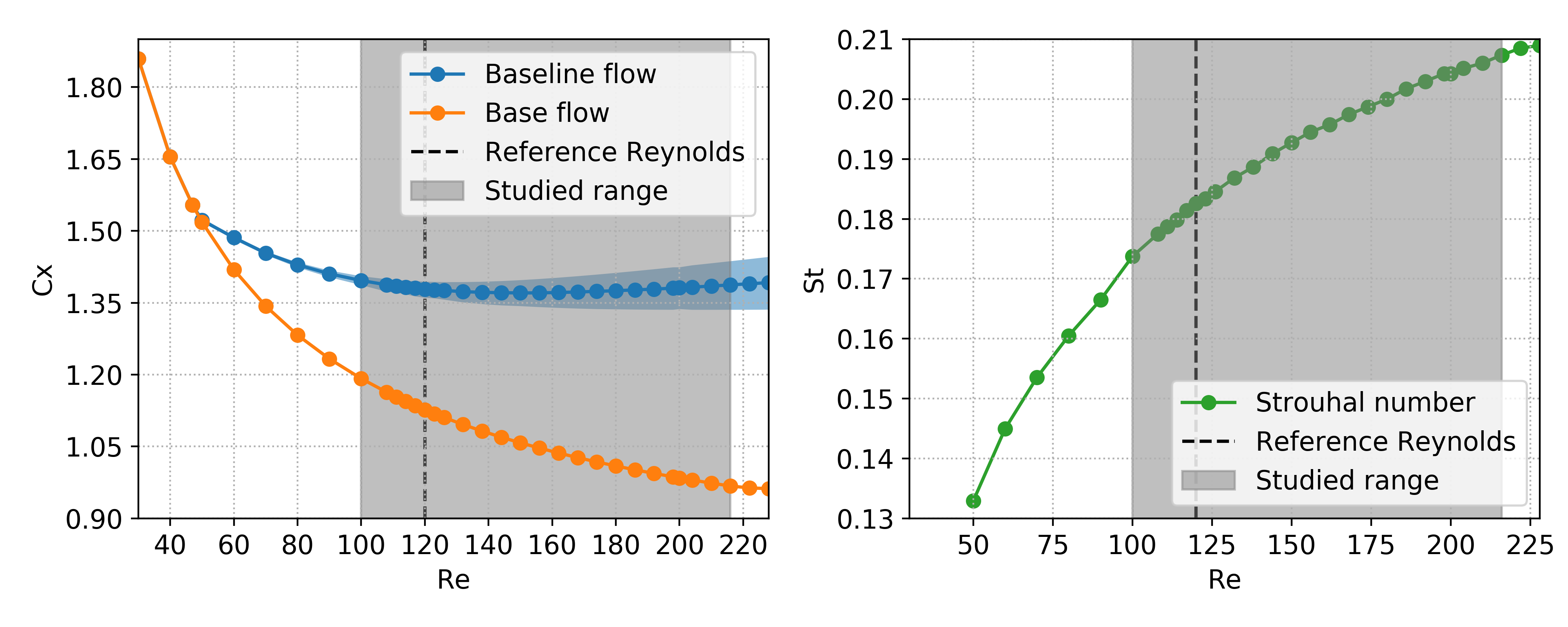}
    }
    \caption{(left): Evolution of the time-averaged drag coefficient of the baseline flow (blue) and the base flow (orange) with the Reynolds number. The blue shaded area indicates the variation range of the drag coefficient $C_x$. (right): Evolution of the Strouhal number of the vortex shedding with the Reynolds number.}
    \label{fig:baselines}
\end{figure}

As presented in figure~\ref{fig:baselines}, values of the drag coefficient and Strouhal number (defined as $St = fD/U_\infty$, with $f$ being the vortex shedding frequency) have been computed for a wide range of Reynolds numbers to ensure consistency with other studies \citep{Nishioka1978,Braza1986,Williamson1996,Henderson1997,He2000,Bergmann2005}. For $\Rey=120$, the drag coefficient is $1.379$ with fluctuations of amplitude $0.018$, and $St = 0.18$, which is in agreement with the literature \citep{Barkley2006,SippMarquetMeligaEtAl2010}. Note that, since the simulation solves the 2D Navier-Stokes equations, the flow remains laminar across the studied Reynolds number range and does not undergo any additional stability bifurcation. 

According to \citet{Protas2002}, the total drag $C_{x,0}$ of the baseline flow can be decomposed into two contributions. The drag of the base flow $C_{x,BF}$, which is constant and the drag correction due to the flow unsteadiness $C_{x,U}$. If $\left<\cdot\right>_{T}$ denotes the time average over a vortex shedding period $T$, then:
\begin{equation}
    \left<C_{x,0}\right>_T = C_{x,BF} + \left<C_{x,U}\right>_T
\end{equation}
Using the base flow performance as a reference, the drag gain $\mu_{C_x}$ which measures the drag reduction due to the control strategy is computed as a fraction of the drag reduction achieved by the base flow:
\begin{equation}
    \mu_{C_x} = \frac{\left<C_{x,0}\right>_T-C_x}{\left<C_{x,U}\right>_T}
\end{equation}
Thus a drag gain $\mu_{C_x}$ of 100\% corresponds to a drag reduction equivalent to a complete suppression of the vortex shedding. \citet{Protas2002} also asked whether a negative mean drag correction $\left<C_x-C_{x,BF}\right>_T$ could be reached with a periodic forcing, implying a drag gain larger than $100\%$. Examples from the literature, such as the work of \citet{He2000} who achieved a drag gain of $108\%$, show that this is possible. But in their case, this performance comes at the cost of a significantly modified mean flow and a large actuation. As shown later, the present study also achieves drag gains slightly higher than 100\%, while both preserving the base flow structure and being energy efficient.

%% file: P2_SPPOCMA.tex
\section{Reinforcement learning algorithms}\label{sec:SPPOCMA}
\subsection{A short description of \textit{on-policy reinforcement learning}}
\begin{figure}
    \centering
    \includegraphics[page=8]{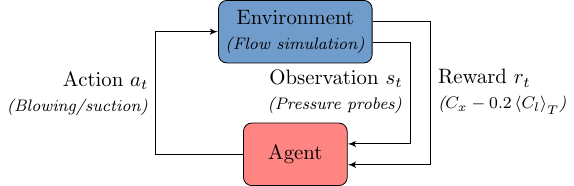}
    \caption{Reinforcement learning feedback loop}
    \label{fig:RL}
\end{figure}

Reinforcement learning considers an environment in interaction with an agent as illustrated in figure \ref{fig:RL}. At each control step $t$, the agent receives partial state observations $s_t$ and a reward $r_t$ quantifying the current performance of the environment. The agent then takes an action $a_t$ based on the observations, through a policy $\pi$: $a_t\sim\pi(\cdot,s_t)$. This policy can either be deterministic or stochastic. The objective of the training is to derive a policy $\pi^*$ that maximises the cumulative reward (called return) throughout time.

In the present study, the environment is the previously described
numerical case and the goal is to minimise drag. Thus the reward is a defined as:
\begin{equation}
    r_t = - C_x - \alpha|\left<C_l\right>_T|
\end{equation}
with $\left<C_l\right>_T$ being a moving average of the lift coefficient on the previous 22 control steps corresponding to the duration of a vortex shedding period, and $\alpha$ the corresponding regularisation coefficient. Here $\alpha=0.2$ (this value is justified in section \ref{sec:perfo}). This penalisation ensures a nearly "zero time-averaged lift" policy. Pressure sensors provide the observed information, actions are computed within a valid range ($a_t\in[-2,2]$) and then implemented on the environment as previously described. The code architecture relies on Tensorflow software library \citep{Abadi2016} and is interfaced with the simulation environment through Cassiopée application programming interface \citep{Benoit2015} using Python programming language. Interface standards are inspired from Open AI's Gym toolkit \citep{Brockman2016}.

All learning algorithms aim at solving the exploration-exploitation dilemma, meaning achieving the best performances at a minimum learning cost. Efficient exploration of the state-action subspace is a key factor in the learning algorithm performance. The exploration is performed through the introduction of randomness in actions. However, too much randomness deteriorates the learning speed and thus reduces exploitation performances for a given learning budget. The careful control of the exploration variance is thus crucial. \textit{On-policy} algorithms try to circumvent this issue using the most recent (and best so far) version of the policy to collect experience, thus sparing inefficient exploration in sub-optimal regions of the state-action subspace.

One of the state-of-art learning approaches is the Proximal Policy Optimisation (PPO) introduced by \citet{SchulmanWolskiDhariwalEtAl2017}. PPO has been successfully used by \citet{RabaultKuchtaJensenEtAl2019,RabaultKuhnle2019,RabaultRenZhangEtAl2020} on a very similar case study. This on-policy actor-critic algorithm uses a dual neural network structure (with around $270,000$ parameters each in our case), an "actor" ($\pi$) and a "critic" ($V$), as agent. Both take the observations $s_t$ as input. The actor outputs an optimal action $\mu_t=\mu_\theta(s_t)$, $\theta$ being the weights and biases of $\pi$. Then, using a predefined standard deviation $\sigma$, an action $a_t\sim \pi_\theta(\cdot|s_t) = \mathcal{N}(\cdot|\mu_\theta,\sigma)$ is sampled, $\mathcal{N}$ being a normal distribution. The critic outputs an estimate of the value $V_t=V_\phi(s_t)$ of the observed state $s_t$, $\phi$ being the weights and biases of $V$. This value is an estimator of the expected return $R_t = \sum_{\tau=t}^{\infty} \gamma^\tau r_\tau$, $\gamma\in]0;1[$ being a discount factor. $V_t$ is used during the update of $\pi_\theta$ to improve learning. The learning phase is performed using a surrogate objective, the updated weights $\theta_{new}$ of the actor aim at making the most successful actions more likely, using the probability ratio $\frac{\pi_{\theta_{new}}(a_t,\mu_t,\sigma)}{\pi_{\theta}(a_t,\mu_t,\sigma)}$. However, to prevent excessively large policy updates, the surrogate objective is clipped. As a consequence, the exploration variance, which is due to both $\sigma$ and the variability of $\mu_t$, often shrinks prematurely as explained by \citet{HaemaelaeinenBabadiMaEtAl2018}. The next section presents a variant of the PPO algorithm, used in this paper, which addresses this limitation. Refer to Appendix \ref{app:PPOCMA} for more detail on PPO.

\subsection{Standard PPO-CMA}\label{sec:PPOCMA}
\begin{figure}
    \centering
    \includegraphics[page=6]{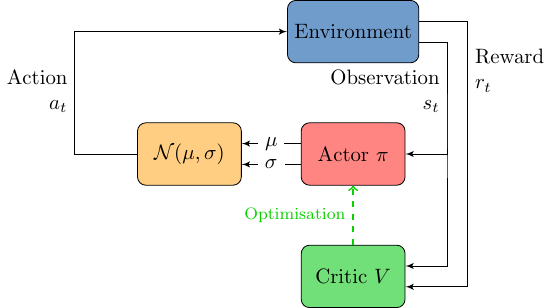}
    \caption{Proximal Policy Optimisation with Covariance Matrix Adaptation (PPO-CMA). The actor $\pi$ receives observations $s_t$ from the environment and outputs $\mu$ and $\sigma$, which are used to sample action $a_t$. The critic $V$ estimates the value $V_t$ of the observed state $s_t$. During the update phase, $V_t$ is used to update the actor and the critic is updated using supervised learning on the effective state values $R_t=\sum_{\tau=t}^{\infty}\gamma^\tau r_\tau$ (observed returns).}
    \label{fig:PPOCMA}
\end{figure}

Proximal Policy Optimisation with Covariance Matrix Adaptation (PPO-CMA) \citep{HaemaelaeinenBabadiMaEtAl2018} is a variant of PPO that prevents the premature vanishing of the exploration variance using the covariance matrix adaptation technique introduced by the CMA-ES evolutionary algorithm \citep{Hansen2003,Hansen2016}.
Unlike PPO, the covariance matrix $\sigma$ used to sample the action $a_t\sim\mathcal{N}(\cdot,\mu,\sigma)$, is an output of the actor $\pi$, as shown in figure \ref{fig:PPOCMA} and the surrogate objective used for updates is not clipped. A more detailed description of PPO-CMA is provided in Appendix \ref{app:PPOCMA}. PPO-CMA is used for all the results introduced in parts \ref{sec:perfo} to \ref{sec:sens_nb_loc}. As shown in the following, it yields significantly improved performances than the "vanilla" PPO algorithm used by \citet{RabaultKuchtaJensenEtAl2019}. 

\subsection{Sparse surrogate actor}
\begin{figure}
    \centering
    \includegraphics[page=7]{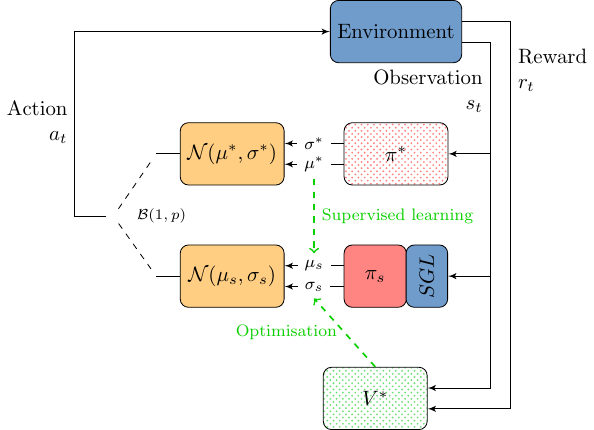}
    \caption{Sparse Proximal Policy Optimisation with Covariance Matrix Adaptation (S-PPO-CMA). Actions are either sampled using the reference actor $\pi^*$ or the sparse actor $\pi_s$ via a Bernoulli choice $\mathcal{B}(1,p)$. $\pi_s$ is updated via learning on $\sigma_s$ using values from $V^*$ and by supervised learning on $\mu_s$ using $\mu^*$ values. The parameters of the stochastic gated layer (SGL) are also updated during this phase.}
    \label{fig:SPPOCMA}
\end{figure}

A novel algorithm called Sparse PPO-CMA (S-PPO-CMA), selecting relevant observations and discarding non-necessary or redundant sensor information, while preserving the optimality of the learned control strategy as much as possible, is introduced. Note that S-PPO-CMA is only used in section \ref{sec:optimalSensors} to optimise the number and location of sensors, the other results presented in this study being obtained with standard PPO-CMA. This method splits into two separate phases: training a conventional PPO-CMA actor-critic structure (described in part \ref{sec:PPOCMA}), then deriving a sparse surrogate actor. The sparse training phase relies on the previously trained PPO-CMA actor-critic policy denoted by $\pi^*$ for the actor and $V^*$ for the critic. As described by figure \ref{fig:SPPOCMA}, the sparse actor $\pi_s$ is composed of a dense neural network having the same structure (architecture and activation functions) as $\pi^*$, to which a stochastic gated input layer (SGL) is added.

During the sparse training phase, the action $a_t$ is either sampled using the optimal policy $\pi^*$ or $\pi_s$, using a Bernoulli random variable. Both training of $\pi^*$ and $V^*$ are stopped, but their outputs are used to train $\pi_s$ and the SGL that make up the sparse version of $\pi^*$.

\begin{figure}
    \centering
    \includegraphics[page=5]{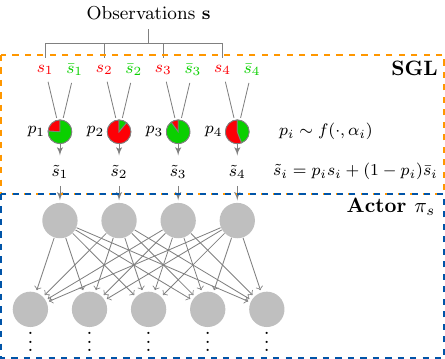}
    \caption{Stochastic Gated input Layer (SGL). Received observation $s_i$ is either passed on to the actor $\pi_s$ if $p_i=1$, combined with its substitute value $\bar{s}_i$ if $p_i\in]0;1[$, or replaced by $\bar{s}_i$ if $p_i=0$. $p_i$ is sampled using a "gating" function $f$ parameterised by $\alpha_i$, which is updated during the second phase the S-PPO-CMA method.}
    \label{fig:SGL}
\end{figure}

The SGL mechanism used here is inspired by the stochastic gate model proposed by \citet{Louizos2017}. Let $n$ be the number of sensors (or the dimension of the observation space). The SGL, presented in figure \ref{fig:SGL} is a special simply connected layer that provides inputs to $\pi_s$ and that contains substitute values $\bar{\vec{s}}=(\bar{s}_1,\bar{s}_2,...,\bar{s}_n)$ for each observation component. Every time an observation vector $\vec{s}=(s_1,s_2,...,s_n)$ is received, the SGL samples a random vector $\vec{p}\in[0;1]^{n}$ that determines its output $\tilde{\vec{s}}$ such as:
\begin{equation}
   \tilde{\vec{s}} = \vec{p}\odot\vec{s}+(1-\vec{p})\odot\bar{\vec{s}}   
\end{equation}
where $\odot$ represents the elementwise product. Thus, $p_i=0$ outputs the observation $s_i$ whereas $p_i=1$ gives its substitute value $\bar{s_i}$, and any value in-between provides a linear combination of $s_i$ and $\bar{s_i}$. Similarly to \citet{Louizos2017}, $\vec{p}$ is sampled over a "gating" function:
\begin{equation}
    \vec{u}\sim \mathcal{U}^n(0,1)
\end{equation}
\begin{equation}
    \vec{p}= f(\vec{u},\vec{\alpha}) =  \text{clip}\left((\zeta-\gamma)\text{Sigmoid}\left[\frac{1}{\beta}(\log \vec{u} - \log (1-\vec{u})+\vec{\alpha})\right]+\gamma,0,1\right)
\end{equation}
with $\mathcal{U}^n(0,1)$ denoting a uniform distribution on $[0,1]^n$, $\beta$, $\gamma$, $\zeta$ being fixed numerical parameters, $\vec{\alpha}$ being a trainable vector steering the expectation on $\vec{p}$ and $\text{clip}(a,b,c) = \min\left(\max\left(a,b\right),c\right)$. $f$ can be seen as a "soft" Bernoulli choice distribution enabling values of $\vec{p}$ in $[0;1]^n$. The $L_0$ complexity of the SGL, giving the expected number of observation components $s_i$ for which $p_i>0$, can be written as:
\begin{equation}
    \mathcal{L}_c\left(\vec{\alpha}\right) = \sum_{i=1}^{n} P(p_i>0) = \sum_{i=1}^n \text{Sigmoid}\left[\alpha_i-\beta\log\frac{-\gamma}{\zeta}\right]
\end{equation}
During testing, $\vec{p}^*$, the most likely value of $\vec{p}$ is chosen deterministically as:
\begin{equation}
    \vec{p}^*=\text{clip}\left((\zeta-\gamma)\text{Sigmoid}\left[\vec{\alpha}\right]+\gamma,0,1\right).
\end{equation}
Both $\vec{p}$ and $\vec{p}^*$ can take values between $0$ and $1$ (included), thus modelling a fully "open" or fully "closed" gate while still allowing for a gradient-based optimisation using the loss $\mathcal{L}_c$.

\subsection{Sparse actor training}
The weights $\theta_s$ of the sparse actor $\pi_s$ are initialised using the weights $\theta^*$ of $\pi^*$ and updated every epoch both by the training of $\sigma_s$ and $\mu_s$. $\sigma_s$ is trained in the same way $\sigma^*$ has been trained (refer to Appendix \ref{app:PPOCMA}). Concerning $\mu_s$ however, a supervised learning using the optimal action $\mu^*$ is performed with the loss:
\begin{equation}
    \mathcal{L}_{\pi_s}\left(\theta_s,\vec{\alpha}\right)=||\mu_s\left(\vec{s},\theta_s,\vec{\alpha}\right)-\mu^*\left(\vec{s},\theta^*\right)||_1
\end{equation}

For the SGL, $\vec{\alpha}$ and $\bar{\vec{s}}$ are trained in the same process as $\theta_s$, allowing $\pi_s$ to "adapt" to the variations of input $\tilde{\vec{s}}$ caused by the updates of the SGL. $\bar{\vec{s}}$ is slowly updated using the observation values $\vec{s}$ at every epoch and updates of $\vec{\alpha}$ are based on the following loss $\mathcal{L}_{\text{sparse}}$:
\begin{equation}
    \mathcal{L}_{\text{sparse}}=\mathcal{L}_{\pi_s}\left(\theta_s,\vec{\alpha}\right)+\lambda\left[\mathcal{H}_{1}\left(\mathcal{L}_c\left(\vec{\alpha}\right)\right)+\mat{\Gamma}\vec{\alpha}\right]
\end{equation}
where $\lambda$ is the regularisation parameter, $\mathcal{H}_1$ is a unitary Huber loss and $\mat{\Gamma}$ can be seen as a Tikhonov matrix that accounts for strong correlations between observations. Its purpose is to penalise $\alpha_i$ whose observation $s_i$ is correlated with any other $s_{j\neq i}$ and thus is redundant (refer to Appendix \ref{app:PPOCMA} for more details). The choice of $\lambda$ drives the equilibrium between sparsity and control performance.

%% file: P3_Results.tex
\section{Results and discussion}\label{sec:results}
\begin{figure}
    \centering
    \includegraphics[width=0.6\textwidth]{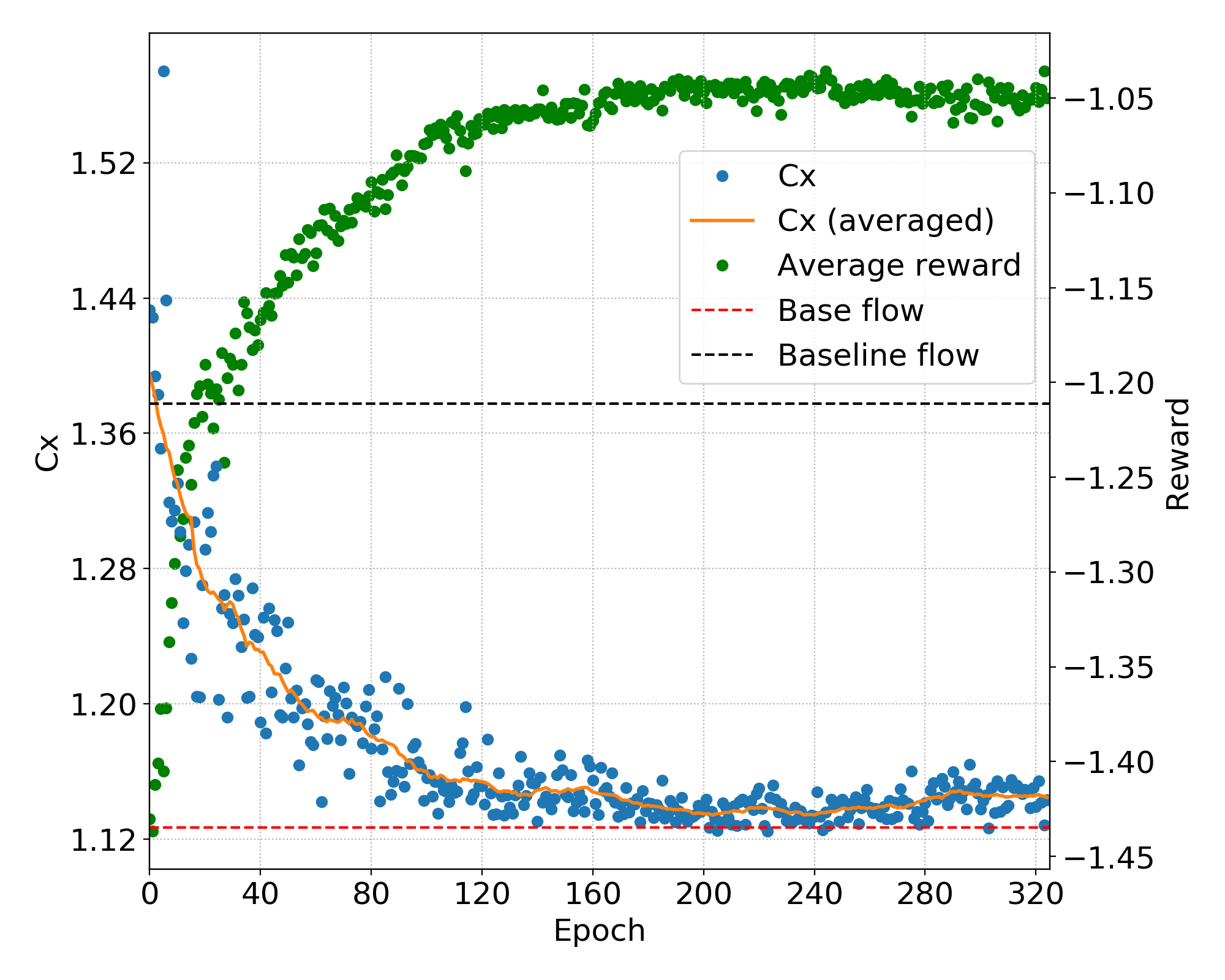}
    \caption{Standard learning process. Each $C_x$ value is averaged over the whole epoch, including the transient from developed vortex shedding to controlled flow. This explains the discrepancy with pure performance values on $C_x$ later introduced. The yellow curve is a 20-epoch moving average of $C_x$ values. The average reward (green dots, reward $r_t$ averaged over the current epoch) shows a quasi-monotonic growth that saturates from epoch $200$ onward.}
    \label{fig:learning}
\end{figure}

Unless otherwise stated, all results are obtained using the reference case at $\Rey=120$, with the 12-sensor layout described by figure \ref{fig:refCase} and PPO-CMA as learning algorithm. Figure \ref{fig:learning} illustrates a standard learning process. A large variation of mean $C_x$ values can be observed in the first epochs of training, then $C_x$ values concentrate more around their moving average. This is caused by PPO-CMA decreasing the exploratory variance $\sigma$ when performance stabilises. For all the following results, training is performed over 200 epochs of 480 steps each. A standard training epoch requires around 180s on 4 CPU cores, most of the CPU time being used to run the environment.

\input{P3A_Perfo}
\input{P3B_Analysis}
\subsection{Robustness}
\input{P3C_Reynolds}
\input{P3D_Noise}
\input{P3E_Sensor_loc}

%% file: P3A_Perfo.tex
\subsection{Control performance and efficiency}\label{sec:perfo}
\begin{figure}
    \centering
    \includegraphics[width=0.8\textwidth]{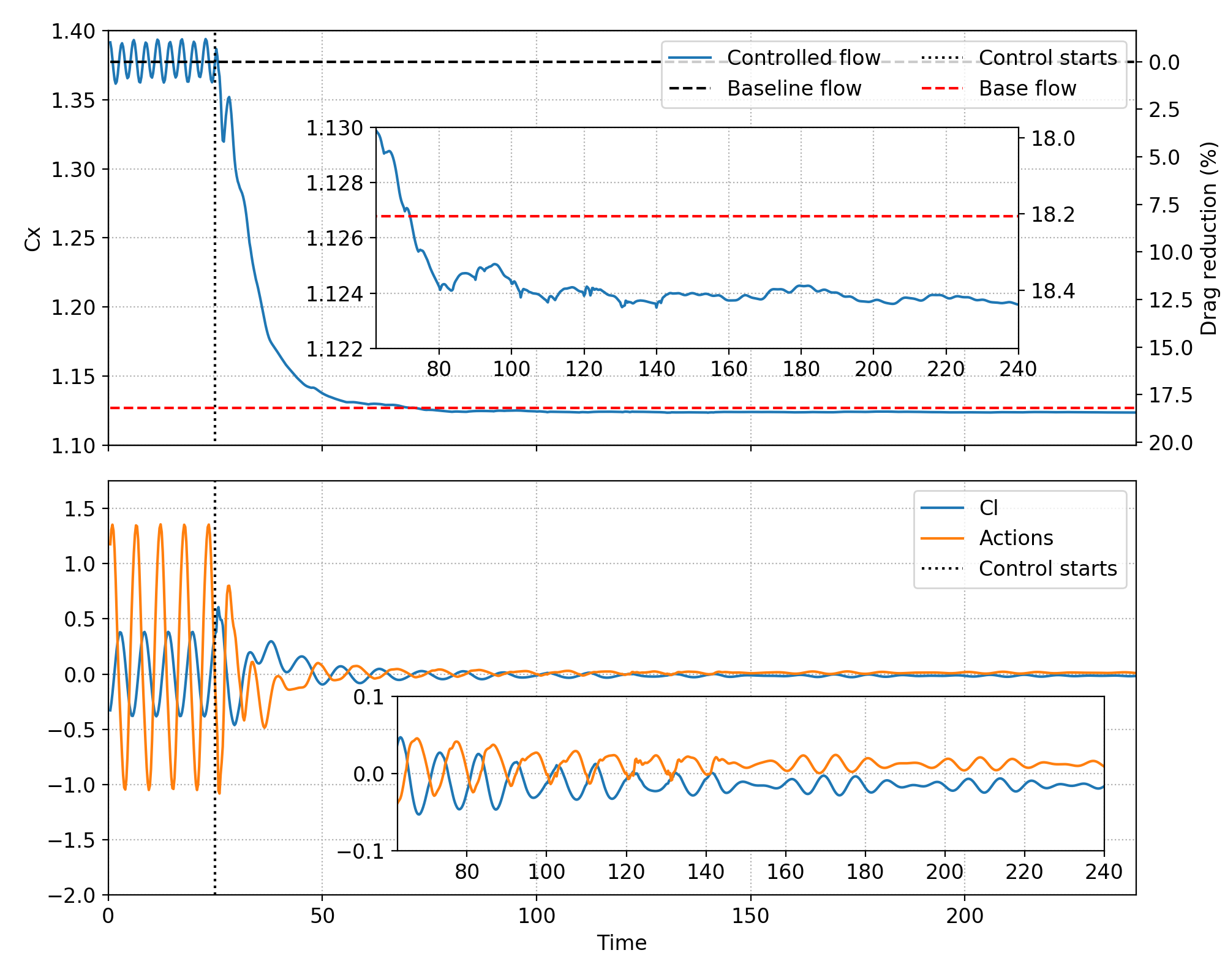}
    \caption{Performance of  the active flow control strategy. (top): Evolution of the instantaneous drag coefficient $C_x$. (bottom): Evolution of both action and lift coefficient $C_l$.}
    \label{fig:ctrl_perfo}
\end{figure}

At a Reynolds number of 120, the time-averaged baseline flow drag coefficient is $\left<C_{x,0}\right> = 1.379$. Performance in terms of drag reduction is computed as a percentage of the average baseline flow drag coefficient $\left<C_{x,0}\right>$ and also using the drag gain $\mu_{C_x}$ introduced in section \ref{sec:unc_flow}. Figure \ref{fig:ctrl_perfo} shows the instantaneous drag coefficient $C_x$, the corresponding action $a_t$ and instantaneous lift coefficient $C_l$ throughout control steps. A first phase, from time $t=25$ (control starting) to $t=50$ approximately, shows a rapid transient from the fully developed vortex shedding instability to the controlled flow. This transient corresponds to approximately $4.5$ vortex shedding periods. During that phase, actions have a large amplitude and do not seem to follow any simple pattern. In a second phase, from time $50$ to the end, the drag coefficient is stabilised to a value below $C_{x,BF}$. This represents a drag reduction of about $18.4\%$ and a drag gain $\mu_{C_x}$ around $100.6\%$. Actions have a significantly reduced amplitude compared to the first phase, and they appear to have a slightly non-zero average. Starting from $t=150$, a periodic action pattern seems to appear in the form of modulated bursts. These last two points are further discussed in section \ref{sec:analysis}.

For other Reynolds number values, drag reduction has also been measured. Results are presented in table \ref{tab:perfo} and confronted to other comparable studies made on the same case. For $\Rey=100$, a drag gain slightly larger than $100\%$ is also reached. The observed mean flow is similar to the base flow. For $\Rey=200$, the results from \citet{He2000} slightly outperform those of the present study in terms of drag reduction. But their drag gain is obtained at the cost of an important mean flow modification, as previously mentioned in section \ref{sec:unc_flow}. Some existing studies on the control of the cylinder flow have not been included in table \ref{tab:perfo} due to the significant discrepancies of their test case compared to ours, which prevents any straightforward comparison. For instance, \citet{Min1999}, using a $360^\circ$ blowing/suction actuation, end up artificially reducing the equivalent diameter of their cylinder, and while their results and methodology are interesting, comparison of performances is however not relevant here. The work of \citet{Arakeri2013}, who impose the tangential velocity on the cylinder surface and force a quasi upstream-downstream symmetric flow, is not included in the comparison for similar reasons. 
Among the related -- yet not directly comparable -- interesting work, we can also cite \citet{Sohankar2015} who achieve a significant drag reduction for a square cylinder flow at $Re=100$, the paper from \citet{Muddada2010} which shows rather important gains using two small rotating rods in the vicinity of the cylinder, or the results from \citet{Chen2005} using magneto-hydrodynamic forcing to stabilise a cylinder flow at $Re=200$. For a more exhaustive list on the topic, one may refer to the review from \citet{Rashidi2016}.

\begin{table}
\begin{center}
\def~{\hphantom{0}}
\begin{tabular}{ccccccc}
 \multirow{2}{*}{Re} & Drag & Drag &\multirow{2}{*}{PSR} & Learning & Action & \multirow{2}{*}{Reference} \\
  & red. ($\%$) & gain ($\%$) &  & type & type & \\
 \hline
\multirow{6}{*}{100} & 8.0 & 54.6 & - & Gradient descent & Blowing & \citet{Leclerc2006}\\
 & 8.0 & 92.7 & - & DRL &Blowing& \citet{RabaultKuchtaJensenEtAl2019}*\\
& 5.7 & 66.1 & - & DRL &Blowing& \citet{Tang2020}*\\
& \multirow{2}{*}{14} & \multirow{2}{*}{95.5} & \multirow{2}{*}{-} & \multirow{2}{*}{ANN/ARX} & \multirow{2}{*}{Translation} & \citet{Siegel2003}\\
& & & & & &\citet{Seidel2009} \\
& \textbf{14.9} & 101.7 & 173 & DRL & Blowing & \textbf{Present study}\\
\hline
\multirow{3}{*}{150}& 4 & 17.5 & - & Parameter study & MHD &\citet{Singha2011}\\
&15 & 65.7 & 51 & Parameter study & Rotation & \citet{Protas2002a} \\
& \textbf{21.2}& 92.9 & 20 & DRL & Blowing& \textbf{Present study}\\
\hline
\multirow{5}{*}{200} &\textbf{31} & 107.9 & - & Adjoint NS &Rotation & \citet{He2000} \\
& 28.6 & 99.6 & 0.07 & POD-based & Rotation & \citet{Bergmann2008} \\
& 24.5 & 85.3 & 0.26 & POD-based & Rotation & \citet{Bergmann2005} \\
& 21.6 & 104.6 & - & DRL &Blowing& \citet{Tang2020}*\\
& 28.6 & 99.6 & 9.2 & DRL &Blowing& \textbf{Present study} \\
\end{tabular}
\caption{\label{tab:perfo} Drag reduction and performance comparison. *These cases are slightly different since walls parallel to the flow are added. Action types: "Blowing": Blowing on cylinder poles, "Translation": Vertical translation of the cylinder, "Rotation": Rotation of the cylinder, "MHD": magneto-hydrodynamic forcing}
\end{center}
\end{table}

\begin{figure}
    \centering
    \includegraphics[width=0.8\textwidth]{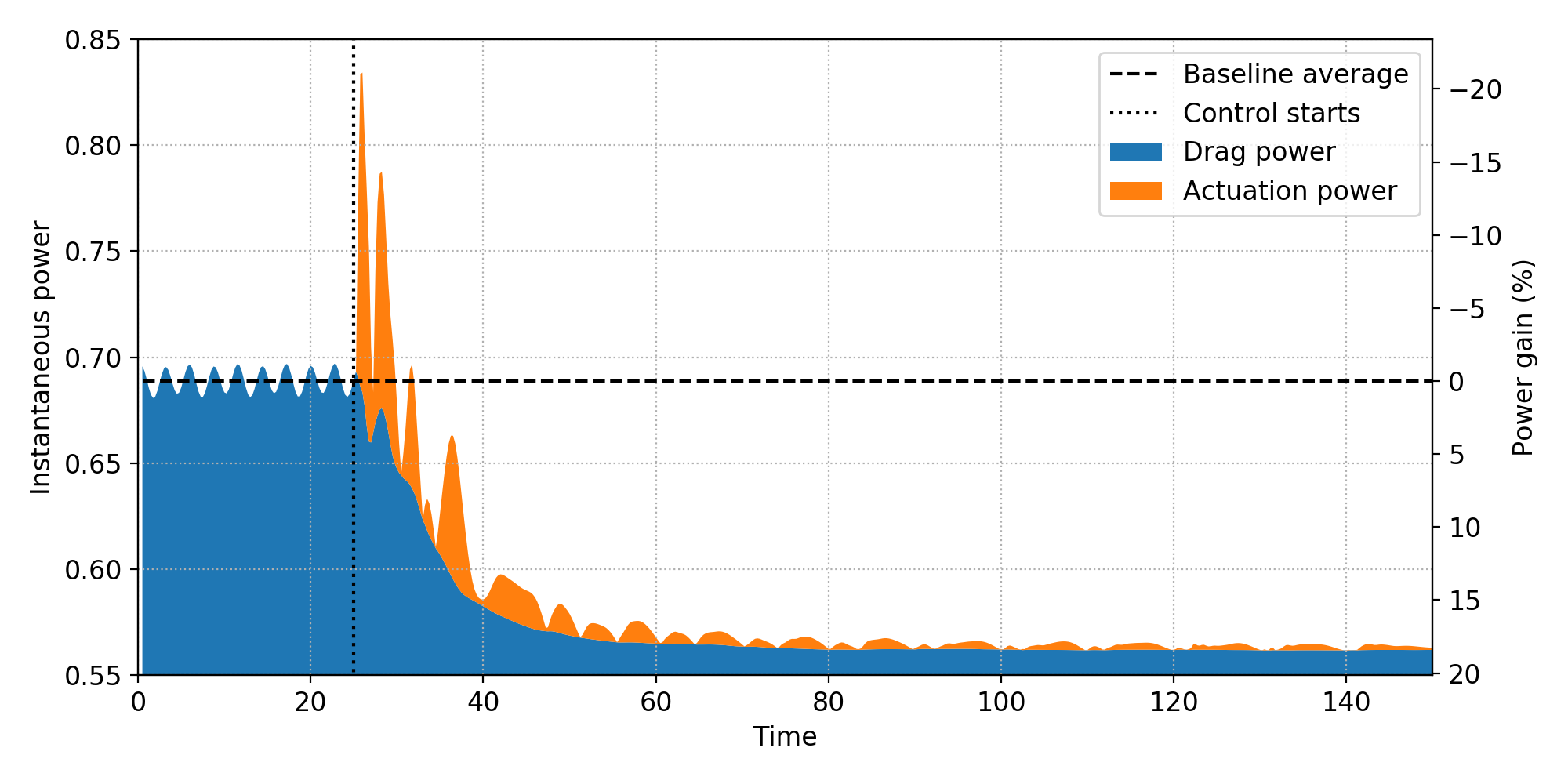}
    \caption{Evolution of power expenditure throughout control. The drag power is the power necessary to withstand drag forces, the actuation power represents the power spent on action implementation.}
    \label{fig:power}
\end{figure}

Another important indicator of the performance of the control is the energy required for drag reduction. Considering the time-averaged baseline flow drag power ($P_{0}=\frac{1}{2}\rho U_\infty^3 D\left<C_{x,0}\right>$) as reference, the actuation power peaks at $22\%$ of $P_0$ in the early stages of the first control phase, but only represents less than $0.3\%$ of $P_0$ on average in the second phase (see figure \ref{fig:power}). Thus the total power expenditure (necessary to both counteract drag and implement action), is temporarily higher than for the baseline flow but is quickly counterbalanced by the significant decrease of both drag and actuation powers during the second control phase. In the example shown in figure \ref{fig:power}, the energy trade-off starts being beneficial $13$ time steps after the control starts, which is long before the flow stabilisation. 

The Power Saving Ratio ($PSR$) was introduced by \citet{Protas2002} and is defined as the ratio of the gain in drag power to the control power. In quasi-steady controlled regime, $PSR\approx71$ for $\Rey=120$, showing that the control obtained here is highly energy-efficient. For other Reynolds numbers, $PSR$ are reported in table \ref{tab:perfo}, and the values found are significantly higher than 1 even for the highest $Re$ considered. 
It can be noticed that for $\Rey=150$, \citet{Protas2002a} achieved extremely energy-efficient control that actually outperform the present study in terms of PSR (but with a lesser net drag reduction). However, their actuation is made through cylinder rotation, and the actuation power does not consider the inertia of the rotating cylinder (the mass of the cylinder is considered null). This highlights that power-based comparisons between different actuation types, especially in the case of cylinder rotation, may have a limited relevance and should be considered carefully.

It is interesting to note that, despite its high energy-efficiency, the control policy is obtained without any explicit penalisation of the instantaneous control power. The actuation power expenditure is not directly included into the measured performance during learning. However, the reward $r_t$ is penalised by the time-averaged lift coefficient, which ensures parsimonious actions since a strong action generates strong lift. Even though $C_l$ is averaged over one vortex shedding period, the periodicity of the flow varies (or even vanishes) during training which makes a perfect compensation of positive and negative actions' effect on $C_l$ very unlikely. As described early on, slightly non-zero-average actions are systematically observed during the second control phase. Thus, a lack of convergence cannot account for this fact. Instead, an increase in the penalisation on lift through $\alpha$ causes a reduction of this constant component.

However, an increase in $\alpha$ has downsides. By trying several values within the range $[0;5]$, it has been observed that, for $\Rey=120$, the chosen value $\alpha=0.2$ is close to the optimal trade-off between pure performance and energy consumption. For both an increase or a decrease of $\alpha$, the PSR decreases and the drag reduction shows a very slight decline. The slight negative effect on the PSR when $\alpha$ decreases below $0.2$ can be explained by the reduction of the penalisation on large actions, thereby increasing the control power expenditure. On the other hand, an overly large value of $\alpha$ reduces the observed exploratory variance due to the strong disadvantage put on large amplitude actions that are necessary in the early stages of the control to achieve a near-stabilisation of the flow.
The search of the optimal $\alpha$ value has only been performed for $\Rey=120$, and the value of $0.2$ has been retained for other Reynolds values. Therefore, the PSR values presented in table \ref{tab:perfo} may not be optimal and might be improved by a careful choice of $\alpha$. But from the results obtained for $\Rey=120$, it appears that $\alpha$ is not a sensitive parameter: it leads to negligible changes in drag reduction performance, and for a wide range of $\alpha$ values, the PSR remains significantly higher than 1.

%% file: P3B_Analysis.tex
\subsection{Analysis of the controlled flow}\label{sec:analysis}
A common difficulty with deep learning approaches is the physical understanding of the results. Unfortunately, no simple action pattern has been noticed throughout the evaluations of the control strategy, whether it is for the first or second control phase. Unsuccessful attempts to reproduce this action behaviour with simpler linear controllers (simple gain and delayed response) might indicate that  complexity is required to reach the observed control efficiency. While it is hard to precisely explain how the control policy acts on the flow to reduce the drag, the present section nonetheless attempts to describe the control based on an \textit{a posteriori} analysis of the flow.

\begin{figure}
    \centering
    \includegraphics[width=\textwidth]{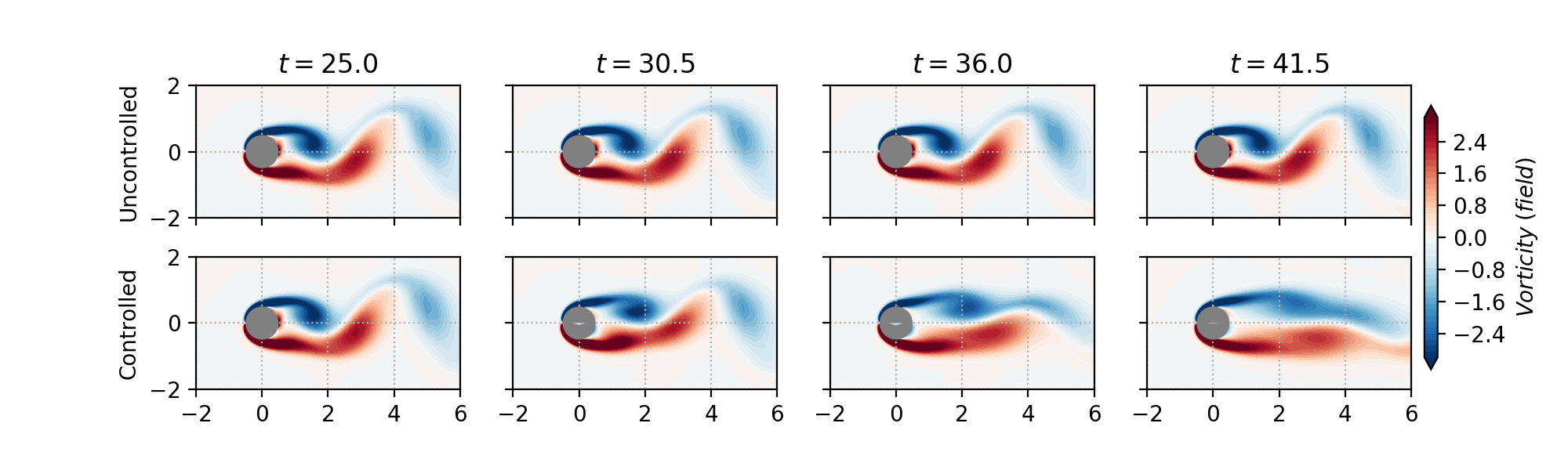}
    \caption{Comparison of uncontrolled (top) and controlled (bottom) flows in the transient phase of the control strategy.}
    \label{fig:transient}
\end{figure}

As studied by \citet{Nair2020}, who used cylinder rotation or momentum injection parallel to the flow to impose an energy optimal phase-shift control, the drag reduction seen in the transient phase, is caused by the delay in vortex shedding. This generates "elongated vortex structures", that also stabilise the instantaneous recirculation bubble. Similar observations were made in our case. As shown by figure \ref{fig:transient}, the first phase of the control strategy is a fast transient from fully developed vortex shedding to a stabilised cylinder wake, where the actions trigger the shedding of vortices slightly earlier than the natural shedding. This results into longitudinally stretched and weaker vortical structures.

\begin{figure}
    \centering
    \includegraphics[width=.8\textwidth]{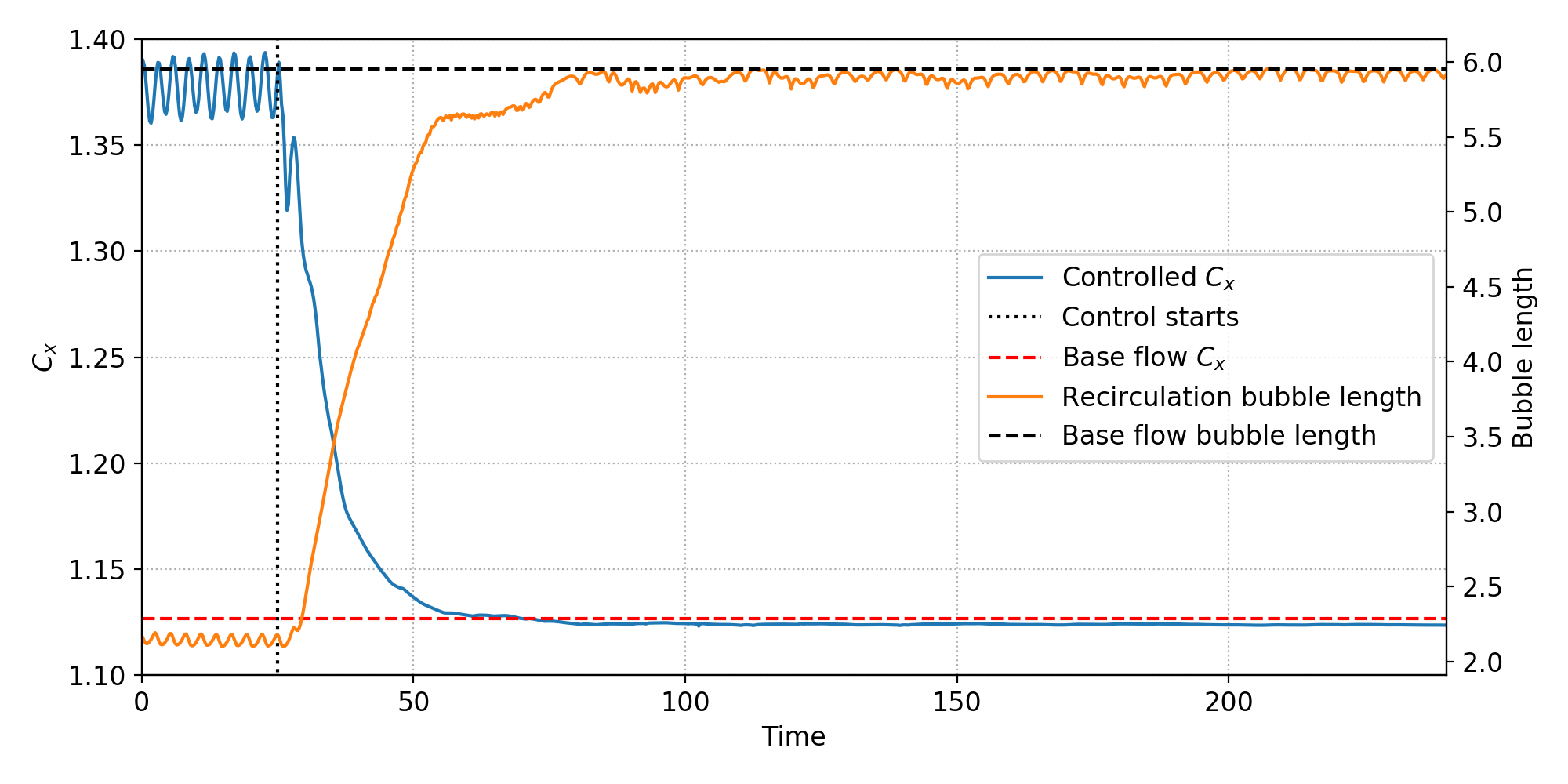}
    \caption{Evolution of $C_x$ and of the length of the instantaneous recirculation bubble throughout time.}
    \label{fig:bubble_perfo}
\end{figure}

\begin{figure}
    \centering
    \includegraphics[width=0.8\textwidth]{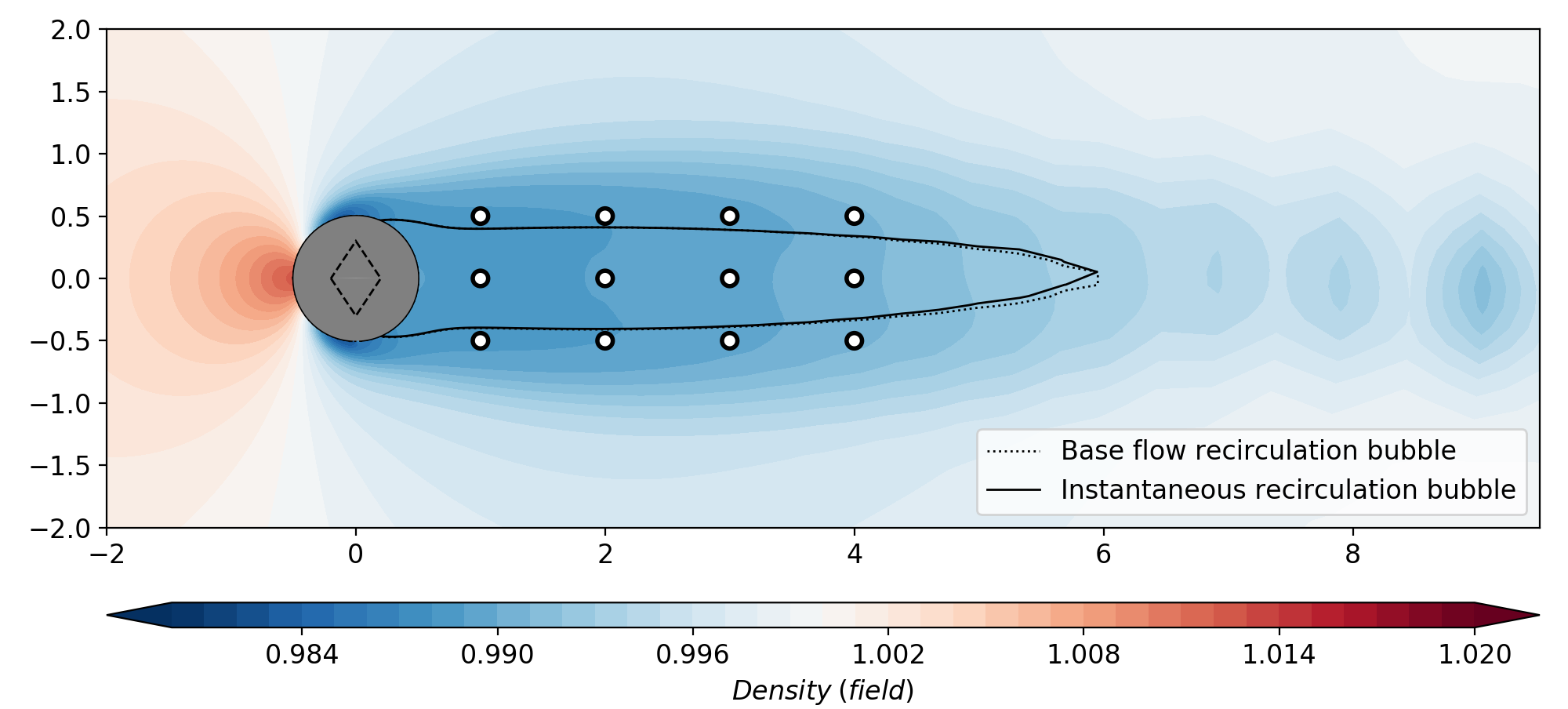}
    \caption{Comparison of the base flow and the controlled flow recirculation bubbles once the flow is stabilised.}
    \label{fig:bubble_flow}
\end{figure}

Once the flow has been stabilised and is nearly steady, its drag coefficient is very close to $C_{x,BF}$. Figure \ref{fig:bubble_perfo} compares the convergence of $C_x$ with the length of the instantaneous recirculation bubble. This length is multiplied by more than $2.5$ during the control phase and peaks at $99.5\%$ of the base flow recirculation bubble length. The correlation of both the increase of the length of the recirculation bubble and the drag reduction is a well-known fact \citep{Protas2002,RabaultKuchtaJensenEtAl2019}. The recirculation bubble lengths found in this study are in good agreement with the reference literature \citep{Zielinska1997,Protas2002}.
From time step 100 onward, both base flow and controlled flow have a very similar recirculation bubble, as illustrated by figure \ref{fig:bubble_flow}. The "tail" of the controlled bubble slowly flaps vertically with a very moderate displacement amplitude ($\Delta y <0.3$) at $St\approx0.12$. This confirms that the control policy tends to lead the flow towards the base flow, the latter being an unstable optimum with respect to drag. The controlled flow reaches a small amplitude cycle around this equilibrium point.

\begin{figure}
    \centering
    \includegraphics[width=0.8\textwidth]{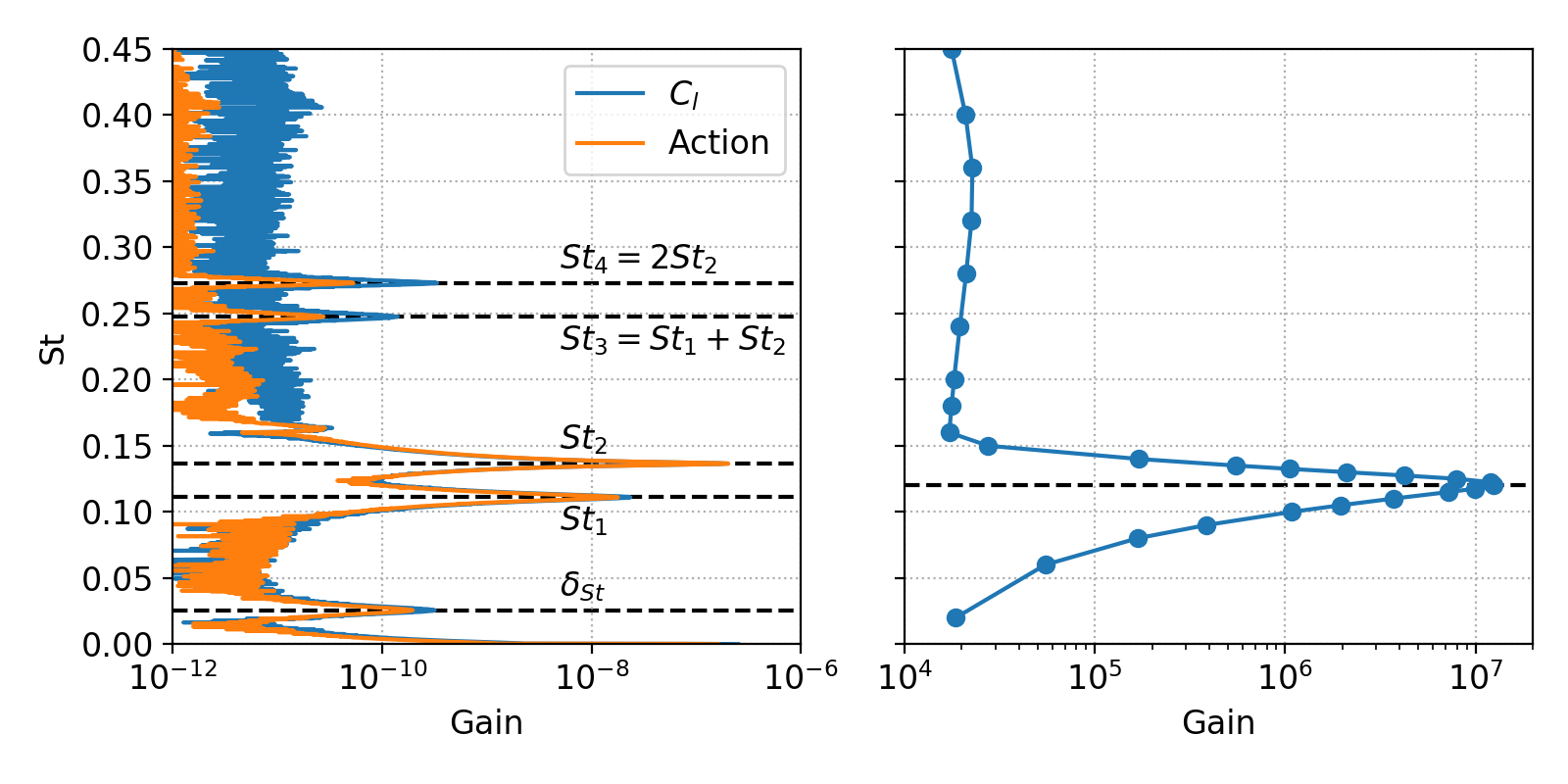}
    \caption{(left): Spectra of the lift coefficient $C_l$ and of the action during the second control phase. $\delta St = St_2-St_1$ (right): Evolution of the optimal gain of the base flow resolvent operator with the external forcing Strouhal number at $\Rey=120$.}
    \label{fig:resolvent_spectra}
\end{figure}

As shown in figure \ref{fig:resolvent_spectra} (left), the spectral analysis of the action during the second phase reveals two main oscillating components $St_1\approx0.11$ and $St_2\approx0.14$ and three secondary peaks at Strouhal numbers $\delta_{St}=St_2-St_1$, $St_3=St_1+St_2$ and $St_4=2St_2$, the latter having amplitudes at least two orders of magnitude lower than the main components. Since $\delta_{S_t}$ and $St_3$ are the marks of nonlinear coupling between the two main components, one can assume a nearly interaction-free superimposition of the two main waves at $St_1$ and $St_2$. Their corresponding Fourier modes (not shown here) peak near the location of the "tail" of the recirculation bubble.

Note that in the stabilised phase, the state is close to the base flow and the actions are small, such that the flow evolves in a linear regime. The dominant Strouhal numbers of this phase are significantly lower than the natural vortex shedding frequency $St=0.18$. This may be easily understood by performing a resolvent analysis, which describes the frequency-response of the flow in the vicinity of a steady state. If $\vec{q}$ denotes the flow state, $\vec{f}$ an external forcing, and $\mathcal{N}$ the Navier-Stokes operator, then Navier-Stokes equations may be written in the compact form:
\begin{equation}
    \pd{\vec{q}}{t} = \mathcal{N}(\vec{q})+\vec{f}.
\end{equation}
Decomposing the flow as the sum of the base flow $\vec{q_{BF}}$ and of a small perturbation $\vec{q'}$ and since the base flow is a steady solution ($\mathcal{N}(\vec{q_{BF}}) = 0$), the fluctuations $\vec{q'}$ are governed by the following first-order approximation:
\begin{equation}
    \pd{\vec{q'}}{t} -\mat{J}\vec{q'} = \vec{f}\quad\text{ with }\mat{J}=\pd{\mathcal{N}}{\vec{q}}(\vec{q_{BF}}).
\end{equation}
The previous equation may then be Fourier decomposed. Denoting the angular frequency by $\omega$, the identity matrix by $\ident$ and the Fourier-transformed variables by a hat notation:
\begin{equation}
    \left(i\omega\ident - \mat{J}\right)\hat{\vec{q}}\vec{'} = \hat{\vec{f}}\quad\rightarrow\quad
    \hat{\vec{q}}\vec{'} = \underbrace{\left(i\omega\ident - \mat{J}\right)^{-1}}_{\mathcal{R}}\hat{\vec{f}},
\end{equation}
with $\mathcal{R}$ being the resolvent operator. The highest singular value of $\mathcal{R}$, which is a function of $\omega$, gives the highest linear gain $\sigma$ that may be achieved through an external forcing (see for instance \citet{Beneddine2017}). Formally, it reads:
\begin{equation}
    \sigma^2(\omega) = \max_{\hat{\vec{f}}} \frac{||\hat{\vec{q}}\vec{'}||_{\vec{q}}}{||\hat{\vec{f}}||_{\vec{f}}},
\end{equation}
with $||\cdot||_{\vec{q}}$ and $||\cdot||_{\vec{f}}$ representing norms on the response and forcing spaces respectively (classically associated with the kinetic energy for the response, and the $L_2$-norm for the forcing). As illustrated by figure \ref{fig:resolvent_spectra} (right), the highest optimal gain is obtained for $St=0.12$ (consistently with \citet{Barkley2006,Jin2019}) and the flow is responsive to only a narrow range of Strouhal numbers (below 0.15). It is therefore not surprising that the values associated with the control fall within this range. But interestingly, the control avoids the highest gain frequency and the particular selection of the two specific frequencies $St_1=0.11$ and $St_2=0.14$ remains an open question. To our knowledge, this is not reminiscent of any existing work related to the linear control of the vortex shedding near the base flow. 

%% file: P3C_Reynolds.tex
\subsubsection{Reynolds robustness}
An assessment of the control policy robustness across a range of Reynolds numbers has been performed. Unlike \citet{Tang2020} who trained their policy on several Reynolds numbers values ($100$, $200$, $300$ and $400$) and evaluated it on a mix of "seen" and "unseen" Reynolds numbers, our policy has been trained on a single Reynolds value $\Rey=120$, evaluations have been performed on a range spanning from $\Rey=100$ to $216$ and compared with cases specifically trained on those Reynolds numbers. As illustrated by figure \ref{fig:baselines}, this range of Reynolds numbers corresponds to a variation in vortex shedding Strouhal number ($St$) of around $18\%$. Moreover, the non-dimensional amplitude of the pressure fluctuation displays a factor 2 between the two extreme $\Rey$ values considered, showing that the dynamics of the flow, although not radically different, is still noticeably altered in this range of Reynolds number, such that the robustness is tested in actual off-design conditions.

\begin{figure}
    \centering
    \includegraphics[width=0.8\textwidth]{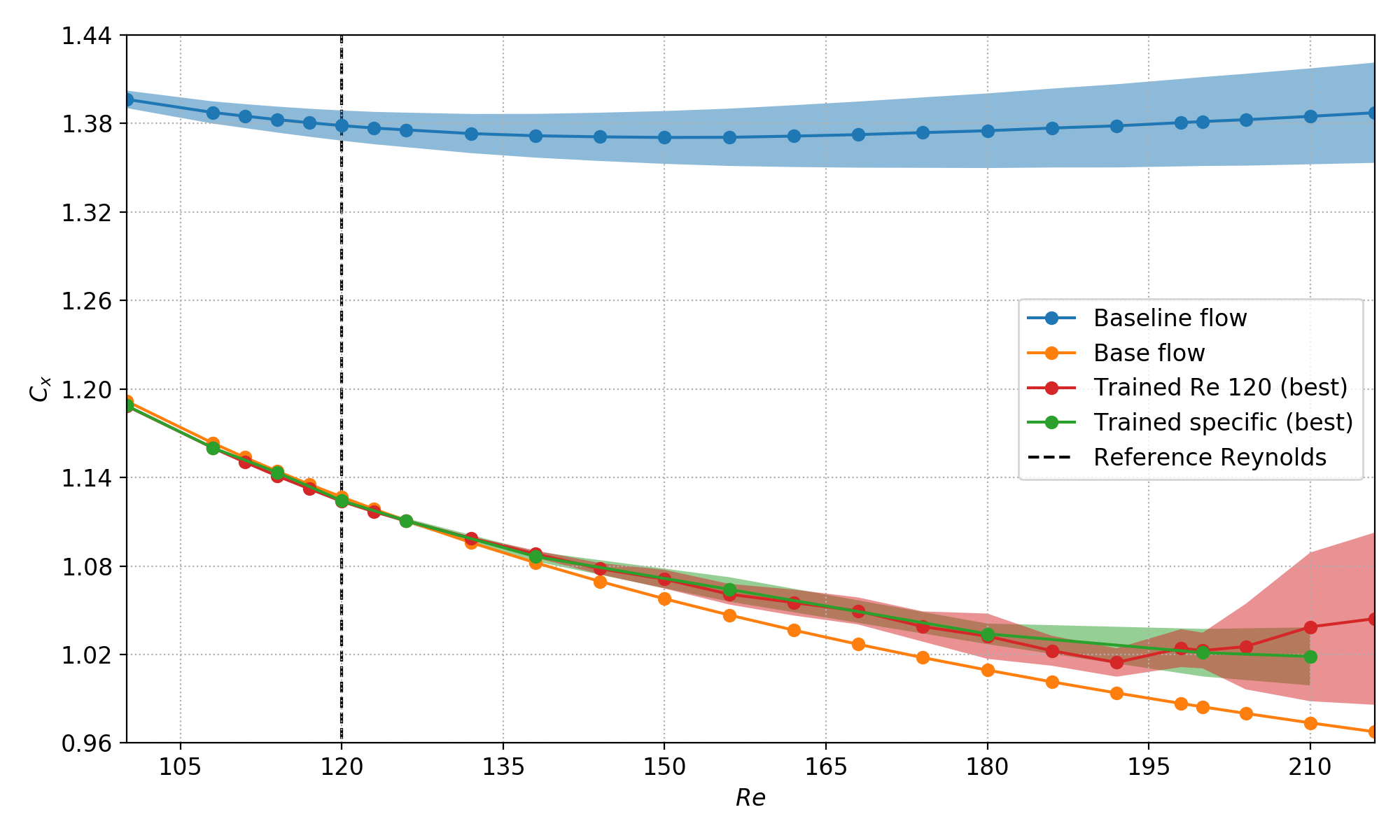}
    \caption{Robustness to a Reynolds number variation. The best case (among 10 test cases) trained at $\Rey=120$ is evaluated at different Reynolds numbers (red curve) and compared to the best control policy (among 10 test cases) specifically trained on the target Reynolds number (green curve). Shaded areas represent the standard deviation of the controlled drag coefficient $C_x$.}
    \label{fig:reynolds_robust}
\end{figure}

Figure \ref{fig:reynolds_robust} shows that the control is remarkably robust. Note that, as previously introduced, the flow state is made non-dimensional by the reference density $\rho_\infty$, upstream velocity $U_\infty$ and static temperature $T_\infty$.
Reference velocities, pressures and case geometry (cylinder diameter and sensor location) being held constant is decisive for the robustness of the control policy. This ensures indeed a nearly constant convection time between sensors and comparable variation amplitudes both for sensors (on pressure) and for actuators (on mass flow) across the different Reynolds numbers considered. The only varying factor between different Reynolds flows is the change in vortex shape, their relative strength and organisation. The policy, acting only as a function of the current observation $s_t$, is insensitive to the variation in the von K\'arm\'an vortex street convection velocity. It is hence only affected by the change in instantaneous form of the flow structures, and the present results proves that the control law handles very well these changes.

Non-dimensonalisation also circumvents the issue of neural network input normalisation. Once neural networks' weights and biases are tuned to adapt to the range of input values, they remain appropriately tuned as this range does not overly change across Reynolds numbers. \citet{Tang2020} used the same non-dimensionalisation scheme. Thus, even though their deep learning algorithm is different, the robustness they observed may be explained by the fact that the policy is robust over a wide range of Reynolds numbers even with a single Reynolds number training. Adding several other Reynolds numbers in the training marginally improves an already strong robustness. 

This robustness is very promising for future experimental exploitation of deep learning for flow control, since flow conditions are subject to uncertainties in experiments. Should one attempt to use a CFD-trained policy to control an actual experimental case, it may be interesting to consider transfer learning, which consists in re-training only specific layers of the network rather than the whole model to quickly adapt it to a slightly different configuration, without restarting from scratch. Several studies have been done in the image processing community on the topic \citep{Huh2016} and should be considered for future work.  

Figure \ref{fig:reynolds_robust} also shows better robustness for lower Reynolds numbers than for higher ones (compared to the training $\Rey$). One of the reasons may be the chosen sensor layout, fixed across all cases but which covers more of the base flow recirculation bubble for lower Reynolds numbers. It has been shown indeed that its length increases with the Reynolds number. The $12$-sensor layout spans over $75\%$ of the recirculation region at $\Rey=100$, but only $55\%$ at $\Rey=216$.

%% file: P3D_Noise.tex
\subsubsection{Observation noise robustness}
\begin{figure}
    \centering
    \includegraphics[width=.8\textwidth]{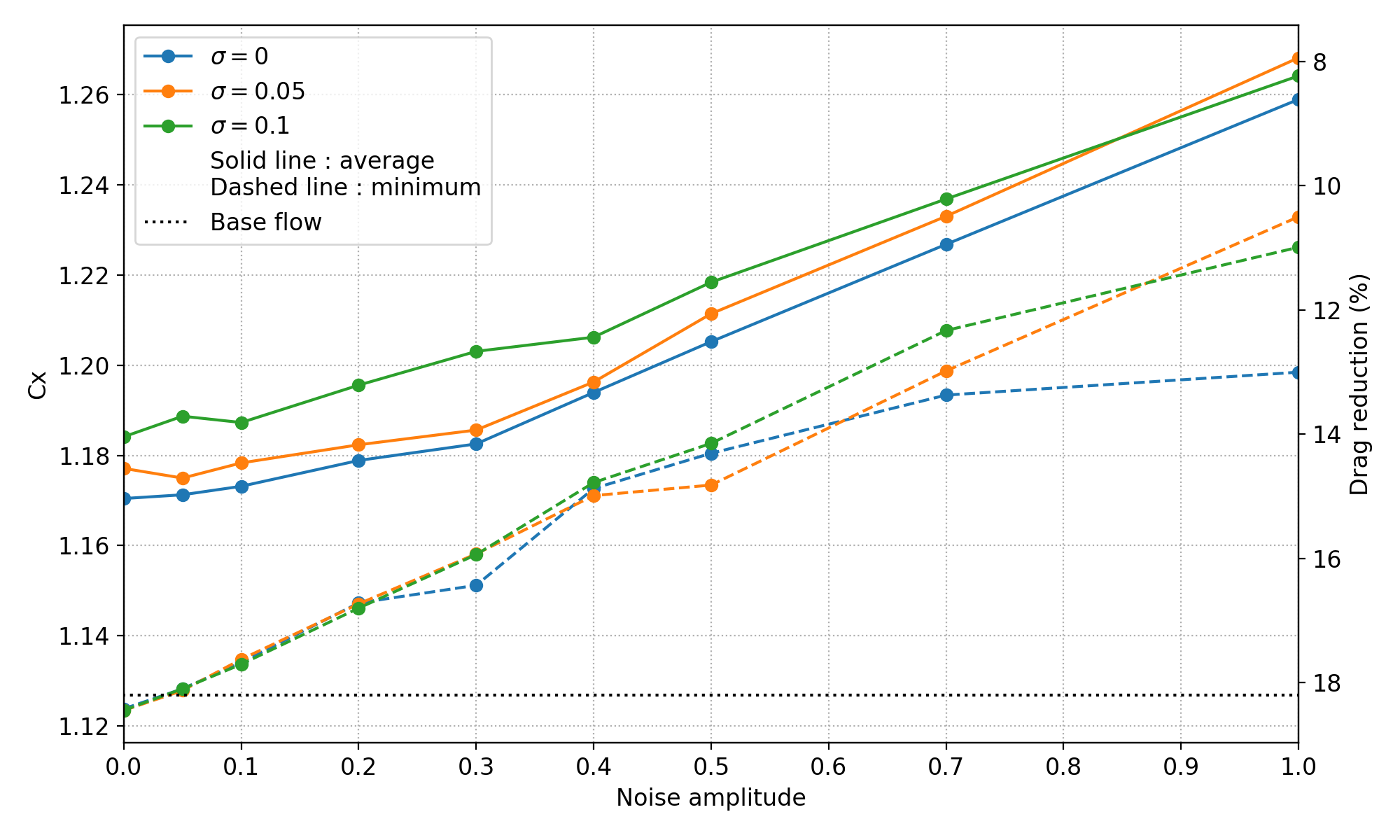}
    \caption{Robustness to Gaussian noise on observations. Each curve represents the mean (solid line) or the best (dashed line) performance in drag coefficient of a 20 test-case batch trained with noise levels from $\sigma=0$ to $\sigma=0.1$ and evaluated on noise levels ranging from $\sigma=0$ to $\sigma=1$.}
    \label{fig:noise_robust}
\end{figure}

\begin{figure}
    \centering
    \includegraphics[width=.8\textwidth]{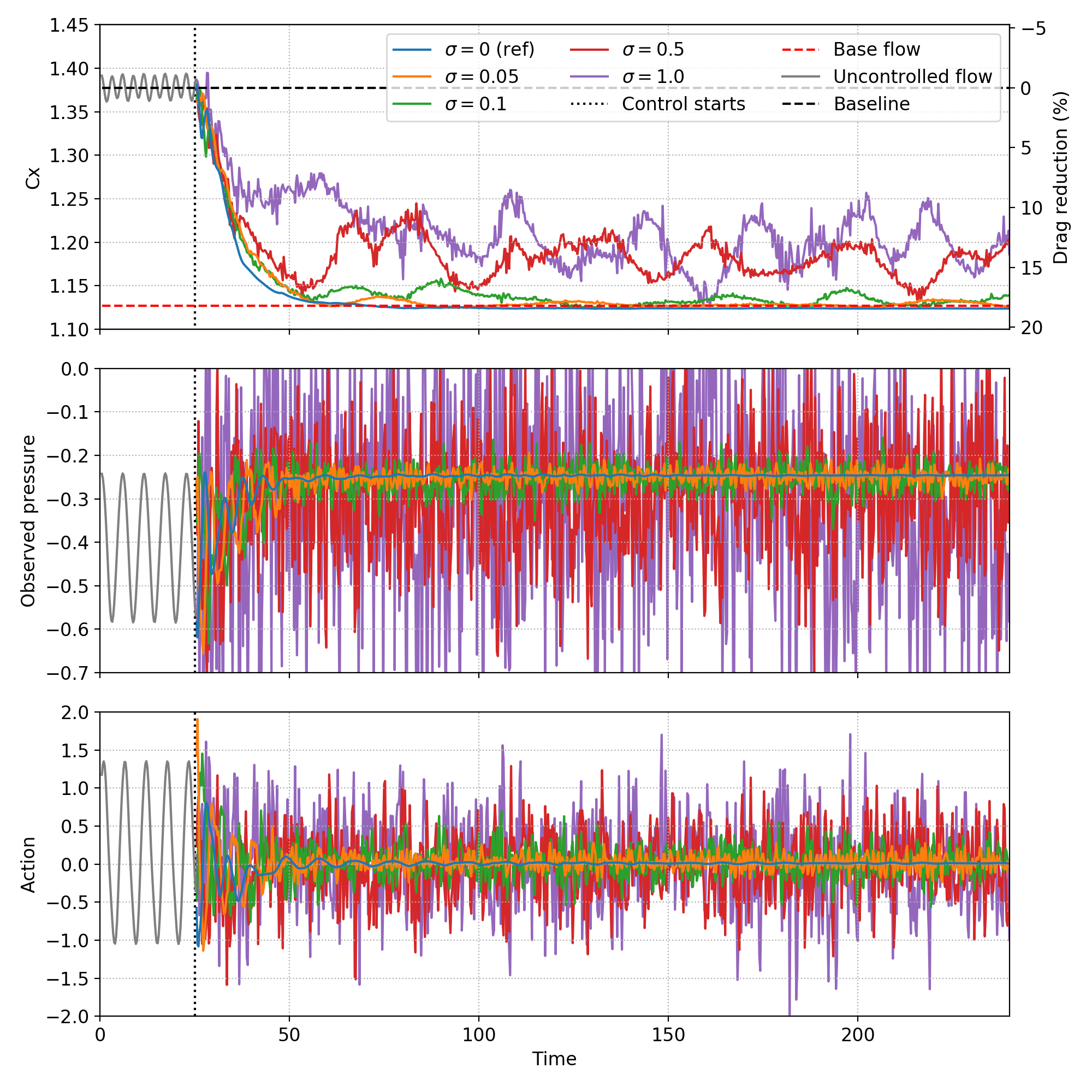}
    \caption{Robustness to Gaussian noise on observations for a policy trained without noise ($\sigma=0$). (top): Evolution of $C_x$ throughout time, for different noise levels. (middle): Noisy pressure signal $s_0$ located in (1,0.5). (bottom): Corresponding action taken by the actor.}
    \label{fig:noise_perfo}
\end{figure}

Assessing the tolerance of the control strategy to measurement errors is a key point in the transposition of that method to real-world experiments, where measurement noise is unavoidable. Noise robustness is therefore important in the perspective of transfer learning from a numerically trained case (without noise) to an experimental setup. To this end, the robustness of a zero-noise-training policy has been assessed and compared with policies trained on noisy data. Added noise is parameterised, using a relative amplitude $\sigma$. Noisy observations $\tilde{s}_t$ are computed as:
\begin{equation}
    \tilde{s}_t = s_t + \bar{s}_t\sigma\mathcal{N}(\cdot|0,1)
\end{equation}
where $\bar{s_t}$ is the average pressure over all sensors at time $t$, which is found to be relatively steady and $\mathcal{N}(\cdot|0,1)$ is a standard random normal probability distribution. Figure \ref{fig:noise_robust} compares the performances of policies trained at different noise levels $\sigma$ and evaluated on a range of noise levels from $0$ to $1$. One can notice that the level of training noise does not seem to impact performances in a significant manner up to $\sigma=0.5$, which corresponds to very noisy measurements that certainly exceeds the actual noise one may expect in most experiments (see figure \ref{fig:noise_perfo}). Unexpectedly, figure \ref{fig:noise_robust}  tends to show that a zero-training-noise policy seems overall slightly more robust to noise than others at different training noise levels. Therefore, in the present case, it is unnecessary to account for measurement noise during the training, which is once again promising for the possible transfer of CFD-trained models to experiments.

Figure \ref{fig:noise_perfo} illustrates this robustness throughout time for a policy trained with $\sigma~=~0$. Despite large noise disturbances, the control policy achieves good performances. Even with extreme noise levels such as $\sigma=1$, the drag reduction reaches about $12\%$ on average. This is only possible in a closed-loop control strategy, and may be explained by the feedback characteristic of the problem that enables for efficient error correction from one control step to the next.
Both observation and action signal-to-noise ratios ($SNR$) are assessed on the second control phase (having steady statistics) as:
\begin{equation}
    SNR=\frac{\text{noise-free variation amplitude}}{\text{noise standard deviation}}
\end{equation}
Action noise is defined as the difference between the action computed using noisy observations and the action based on noise-free measurements.
Results are reported in Table \ref{tab:noise}. Note that apparent discrepancy between $SNR$ and $\sigma$ values are due to the definition of each quantity: $SNR$s are computed considering the amplitude of variation of the signal, thus excluding the signal's time-averaged value, while the noise level driven by $\sigma$ is measured as a fraction of the signal value, including its constant component. It can be seen that the action $SNR$ has the same order of magnitude as the observation $SNR$, their ratio ranges between $0.7$ and $2$. In particular, the $SNR$ of observations and actions become closer as the level of noise increases. This highlights the robustness of the policy, which does not diverge from optimal actions due to spurious fluctuations within the observations. The errors do not accumulate over time and the closed-loop system appears able to rectify the previous erroneous action to contain the deviation from the optimal controlled flow-state. The policy is therefore sufficiently insensitive to input errors in that range of noise levels to ensure a strong robustness. In addition, it is possible that the decorrelation of these errors between each measurements helps mitigating the effects of the noise.
\begin{table}
\centering
\begin{tabular}{ccccc}
 $\sigma$ & Observation SNR & Action SNR & Average drag reduction $(\%)$\\[3pt]
0 & $\infty$ & $\infty$ & 18.4\\
0.05 & 0.33 & 0.17 & 18.1\\
0.1 & 0.18 & 0.09 & 17.8\\
0.5 & 0.04 & 0.04 & 14.2\\
1 & 0.02 & 0.03 & 13.1\\
\end{tabular}
\caption{\label{tab:noise} Noise robustness comparison. SNR: Signal-to-Noise Ratio}
\end{table}

%% file: P3E_Sensor_loc.tex
\subsection{Impact of the sensor number and location on the control performance}\label{sec:sens_nb_loc}
\begin{figure}
    \centering
    \includegraphics[width=.8\textwidth]{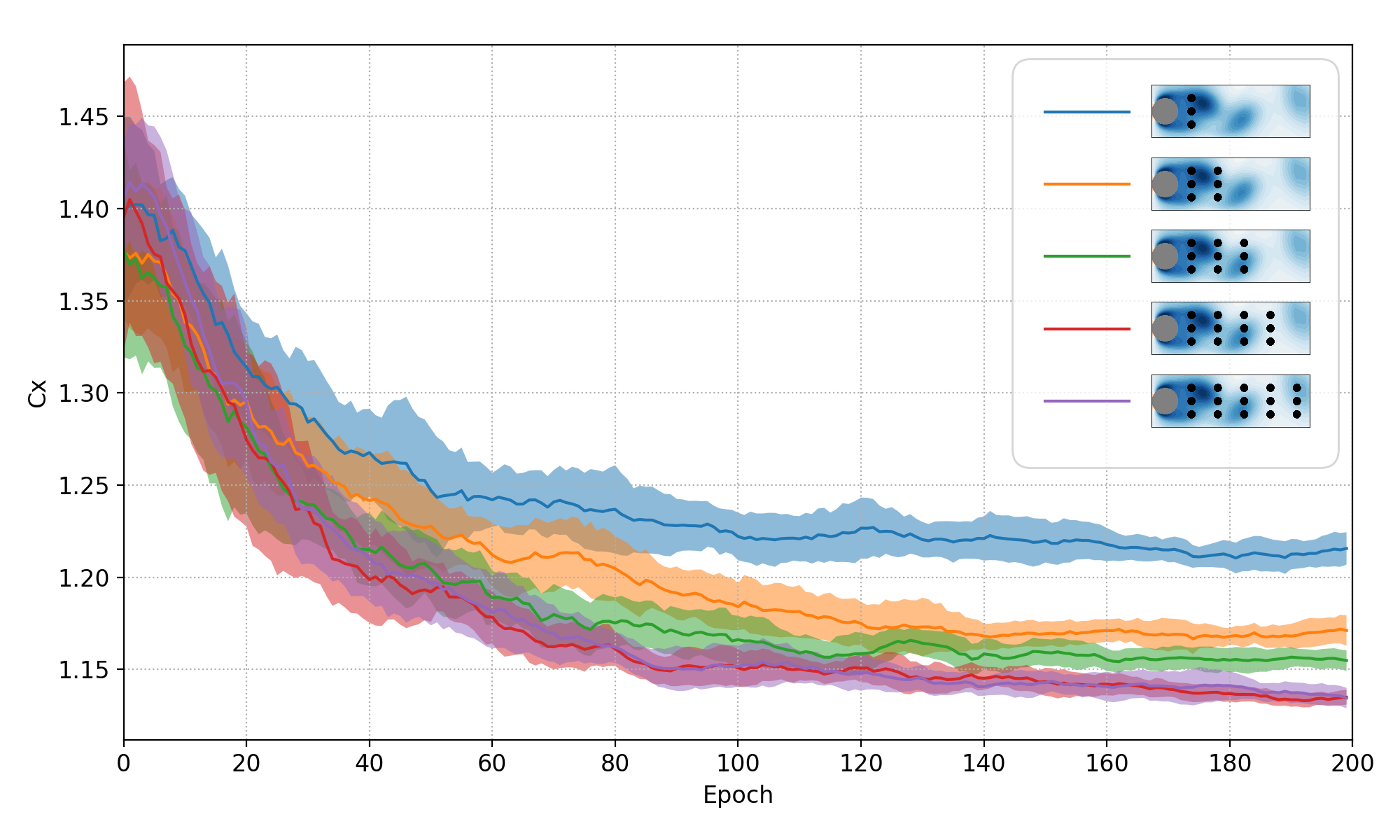}
    \caption{10-case batch-averaged learning curves for different sensor layouts. Shaded areas represent the standard deviation of the corresponding plotted quantities.}
    \label{fig:sensor_nb}
\end{figure}

As introduced previously, the optimisation of both the sensors number and location is a widely explored domain. In this part, a systematic study on sensor configurations within a 3-by-5 grid-like layout is performed. Figure \ref{fig:sensor_nb} illustrates the learning curves of 10-case batches having from $3$ to $15$ sensors. The addition of the second and third columns of sensors (located in $x=2$ and $x=3$) yields a significant gain in performance, and one can notice that $12$ and $15$-sensor layouts have a very similar average performance. Thus it is possible to conclude that the three additional sensors (located in $x=5$) are not useful to the control strategy.

\begin{figure}
    \centering
    \includegraphics[width=.8\textwidth]{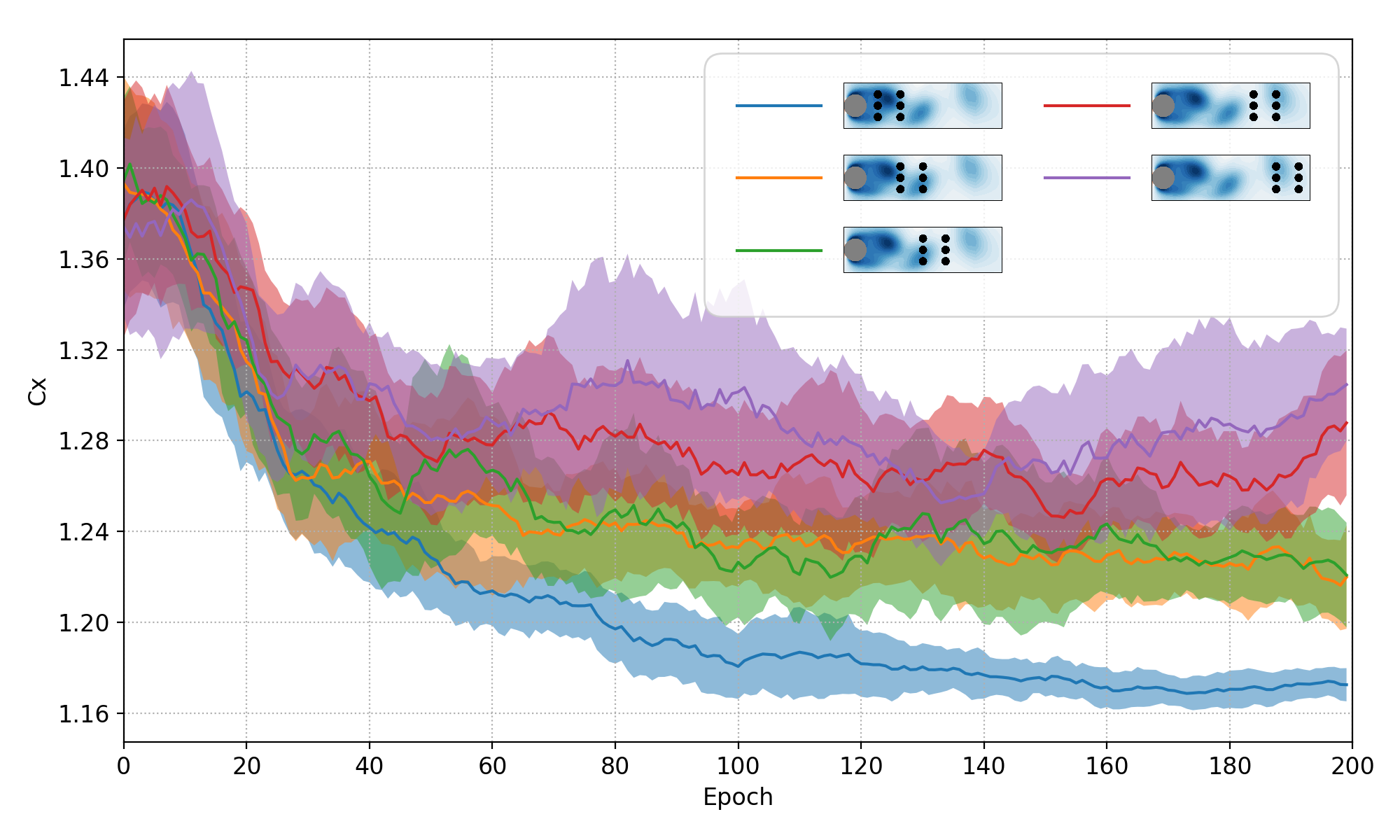}
    \caption{10-case batch-averaged learning curves for different sensor layouts. Shaded areas represent the standard deviation of the corresponding plotted quantities.}
    \label{fig:sensor_loc}
\end{figure}

Figure \ref{fig:sensor_loc} shows the effect of the location of pressure observations, for an array of 6 sensors that are displaced in the streamwise direction. This time, the importance of the first sensor column (located in $x=1$) is demonstrated by the noticeable gain in drag reduction between the first two layouts (blue and yellow curves). The importance of the third sensor column ($x=3$) is once again stressed by the decrease in performance between the green and red curves. Within this predefined combinatorial set, this partial study highlights the relevance of sensors closest to the cylinder. These first preliminary tendencies are confronted in the next section with the results from the newly-proposed S-PPO-CMA algorithm that is designed to provide the optimal sensor location for the control.
 
\subsection{Optimal choice of sensors: results from the S-PPO-CMA algorithm}\label{sec:optimalSensors}
\begin{figure}
    \centering
    \includegraphics[width=.8\textwidth]{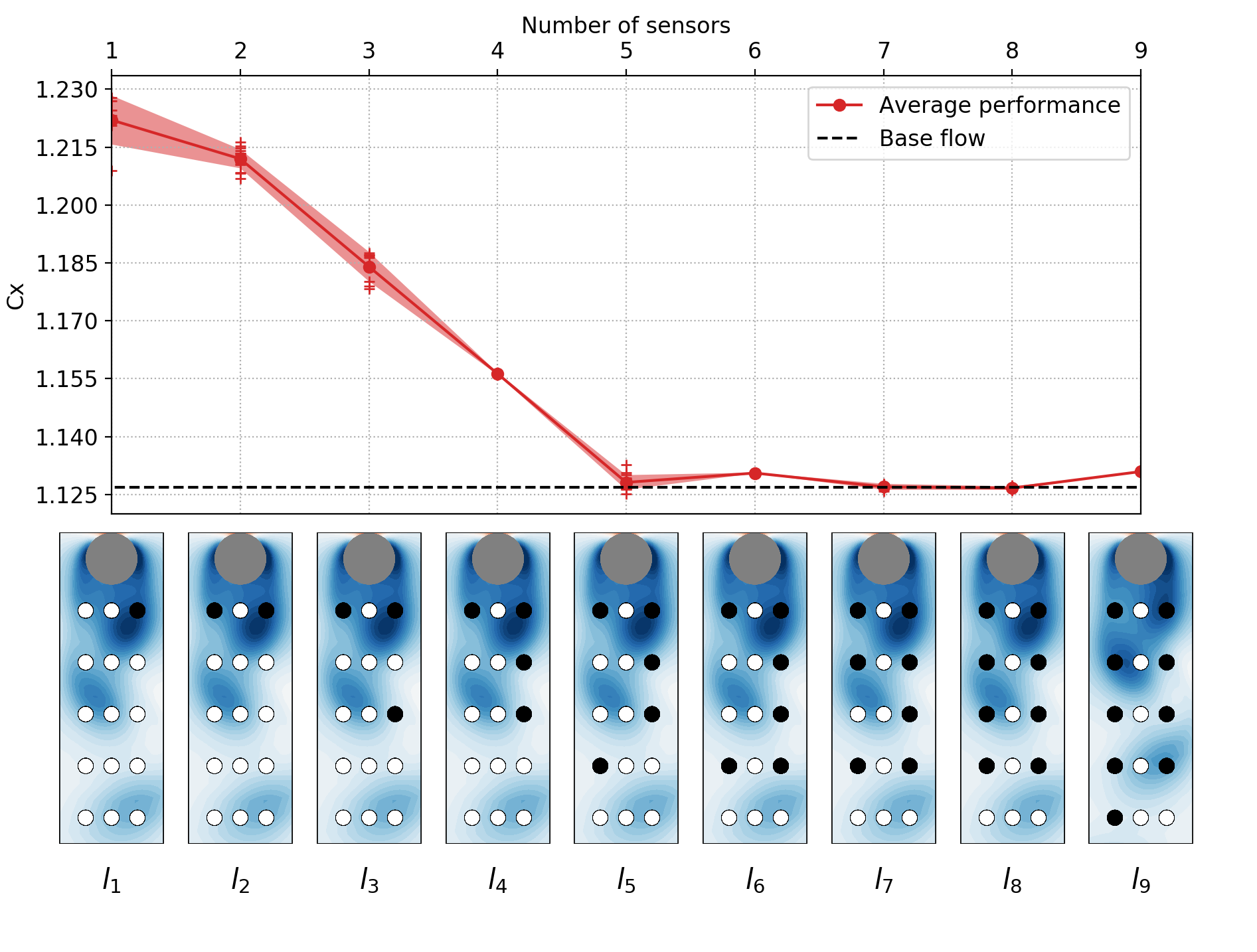}
    \caption{Evolution of both drag reduction and optimal sensor layout with the number of sensors. Layout thumbnails $l_1$ to $l_9$ are rotated $90^\circ$ clockwise.}
    \label{fig:sparse_layout}
\end{figure}

The S-PPO-CMA algorithm, described in section \ref{sec:SPPOCMA}, is used here to derive optimal sensor placement for any allocated number of sensors ranging from $1$ to $9$, within the imposed $15$-sensor grid-like pattern. The number of sensors is indirectly controlled through the value of the $L_0$ regularisation constant $\lambda$, that balances the gradients of both $\mathcal{L}_{\pi_s}$ (performance loss) and $\mathcal{L}_c$ (complexity loss). Figure \ref{fig:sparse_layout} shows the achievable drag reduction with respect to the number of sensors $i$ and the corresponding sensor layout $l_i$. Note that due to the symmetry of the configuration, there always exist pairs of symmetric layout that achieve identical performances. The S-PPO-CMA algorithm randomly outputs one of the two optimal layouts for each value of $\lambda$, but only one layout is displayed in the figures for simplicity. 

With a single sensor, the drag reduction is around $11\%$ and it peaks to approximately $18\%$ for 5 sensors or more. The sensor pattern's tendency to fill without relocating existing sensors, meaning that $l_i\subset l_{j>i}$, is a sign of convexity of this problem in the sense that any combination of the optimal layout set is also part of this set. It is interesting to notice that, starting from the 5-sensor optimal layout, the addition of more sensors does not improve drag reduction, which makes this 5-sensor layout the optimal trade-off between performance and sensor setup complexity. To our knowledge, this layout is not reminiscent of anything used in the large number of existing studies on the control of the 2D cylinder wake. Thus, this highlights the usefulness of the S-PPO-CMA algorithm since optimal sensor placement is, even in such a simple case, not particularly intuitive. 

The centerline locations ($y=0$) do not appear relevant for the control since the corresponding sensors are never selected by the algorithm. A possible explanation may be that these sensors cannot provide information relative of the instantaneous asymmetry of flow, and are thus not fit to choose the action's sign. The first two layouts $l_1$ and $l_2$ validate the importance of the first sensor column, and the selection of sensors shows a weaker importance of locations beyond $x=4$. This is in line with the conclusions of section \ref{sec:sens_nb_loc}.

\begin{figure}
    \centering
    \includegraphics[width=\textwidth]{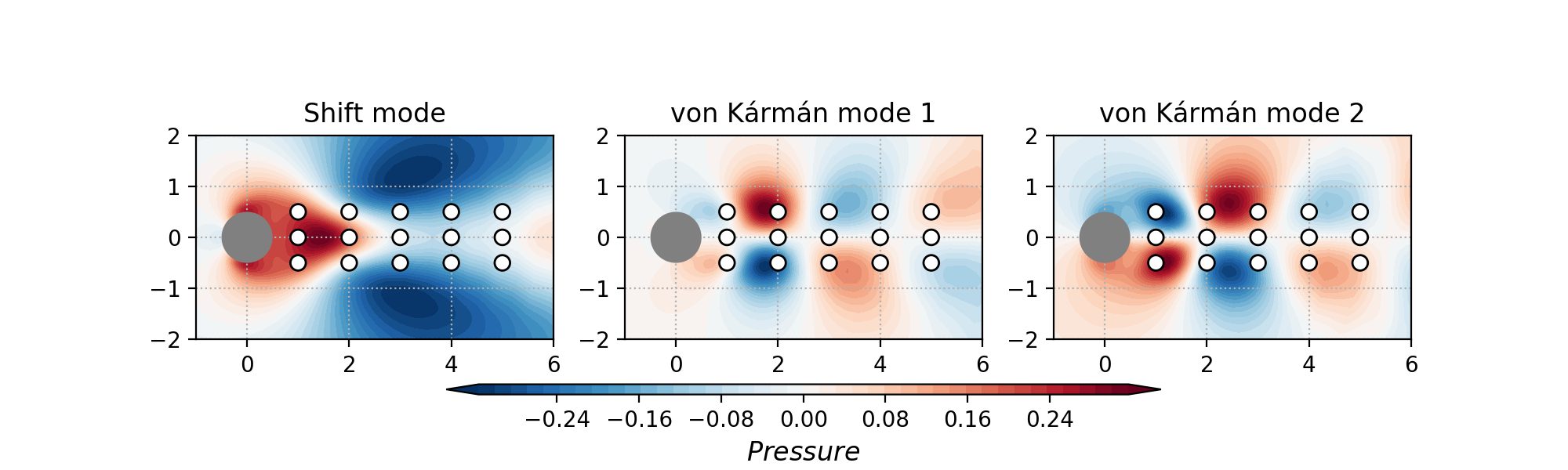}
    \caption{Comparison of the sensor locations with the first three POD modes of a natural transient from base flow to fully developed vortex shedding.}
    \label{fig:POD_layout}
\end{figure}

As discussed in the introduction, many studies optimise sensor placement based on the linear framework of POD, with the underlying idea that the better the estimation of mode coefficients is, the better the reconstruction and control performance are. They naturally often choose locations where the POD modes are strong. Figure \ref{fig:POD_layout} illustrates the superimposition of the sensor locations with the first three POD modes derived from the natural transient from base flow to fully developed vortex shedding. These three modes account for more than $95\%$ of the transient's energy. Despite that these modes are only valid for control trajectories that stay close to this natural transient, the choice of $l_2$ seems reasonable as it allows estimations of both shift mode and second vortex shedding mode simultaneously, since sensors are close to the extrema of this modes (refer to left and right panels of figure \ref{fig:POD_layout}). The second column of sensors appears less able to provide relevant information on the shift mode. Figure \ref{fig:POD_layout} also confirms that the centerline sensors are unfit to estimate von K\'arm\'an modes, which account for the instantaneous asymmetry of the vortex shedding.

Comparing the second and third layouts $l_2$ and $l_3$, it appears that, given the first two probe locations, an additional sensor is preferred in the third column rather than in the second. This might be because this layout provides a better "coverage" of the instantaneous recirculation bubble. Additionally, despite the lack of mean flow symmetry during the transient phase of control, sensors tend to concentrate on a single streamwise row. From a POD-based control viewpoint, this may not be optimal for modes reconstruction. This shows a strength of the present approach: an estimation of the full flow field (or the dominant POD modes) is likely not to be needed for the control. Therefore, searching for points that allow such a full reconstruction may be sub-optimal. In that context, favouring a precise vortex tracking, whose centre travels in the vicinity of the external sensor rows over a more complete estimation of the wake may lead to better performance. 

\begin{figure}
    \centering
    \includegraphics[width=.8\textwidth]{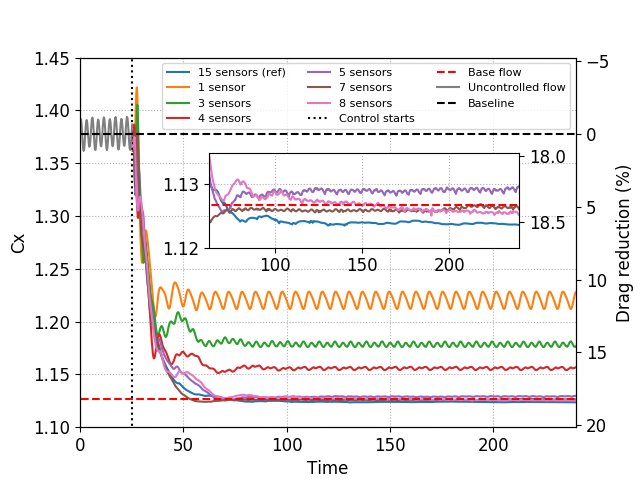}
    \caption{Performance comparison for different optimal sensor layouts.}
    \label{fig:sparse_perfo}
\end{figure}

Figure \ref{fig:sparse_perfo} compares the performances of some the previously found sensor layouts throughout time. It confirms that as from 5 sensors, optimal performance ($\sim 18\%$) is reached. All configurations show a comparable transient phase, that stops earlier both in time and drag reduction for the sparsest sensor configurations. Despite an already significant drag reduction, layout $l_1$ seems unable to notably stabilise the vortex shedding instability. A much better steadiness is achieved with $3$, $4$ and more sensors.

%% file: P4_Conclusion.tex
\section{Conclusion}
In this study, PPO-CMA, implemented on a laminar 2D cylinder case study, discovered efficient nonlinear control strategies with only 12 pressure sensors in the cylinder's wake. A drag reduction of $18.4\%$ is reached for $\Rey=120$. Comparable performances are achieved for other values of $\Rey$, which match state-of-the-art control performances. The control strategy has been analysed \textit{a posteriori} and split into two distinct phases. After a rapid transient phase where large amplitude actions bring the mean flow close to the base flow, the policy simply keeps the controlled flow on a small amplitude limit cycle with weaker actions, whose temporal spectrum is dominated by two distinct frequencies, both close to the base flow's resonance frequency. This translates to a temporary disadvantageous energy expenditure, that quickly becomes beneficial with a $PSR$ in the order of $\textit{O}(10)$ for the whole range of $\Rey$ considered.

The robustness with respect to a variation in Reynolds number and measurement noise has also been quantified. It has been demonstrated that a policy trained at $\Rey=120$ shows near-optimal performances that match the drag reductions achieved by specifically trained policies on Reynolds numbers in the range $[100;216]$. Such robustness is in part explained by the chosen non-dimensionalisation scheme of the case study. The impact of measurement noise has been assessed by both training policies with different noise levels and by evaluation of these policies on different levels of noisy observations. It has been concluded that overall, the control is very robust, yet noise-free trained policies seem slightly more robust than those trained on noisy data and that the actor takes advantage of both the decorrelation of noise between observations and the closed-loop nature of the problem to demonstrate efficient control on noisy environment. Those two aspects show interesting possibilities for direct application of reinforcement-based feedback control on real cases and, in the scope of transfer learning, for a synergistic coupling of numerical simulation and experiments for active flow control.

An optimisation of the number of sensors has also been performed while preserving performances. After a first coarse systematic study showing the importance of having sensors close the cylinder, S-PPO-CMA has been introduced and implemented on the test case. This new algorithm, discovering optimal sensor layouts for reinforcement-based control, selects the most relevant sensors and discards redundant and irrelevant ones. Thus, the number of sensors has been reduced to only 5 while keeping state-of-the-art performance. The obtained sensor layout has been compared with both the outcomes of our systematic study and conclusions of other linear (mostly POD-based) studies. Several explanations have been proposed to back the observed consistency of these results.

A future study could consider extending this approach to larger sensors layouts for more complex cases. One could try to improve performance on the present case study. As shown by \citet{He2000} this could mean seeking for other mean flow configurations with for instance induction of a reverse von K\'arm\'an street \citep{Bergmann2006}. 
This would more likely require a system similar to what \citet{Tang2020} did and a much lower energy efficiency. 

\section*{Acknowledgement}
This work is funded by the French Agency for Innovation and Defence (AID) via a PhD scholarship. Their support is gratefully acknowledged. The authors would like to thank Jean Rabault for the valuable discussions and advice.
\section*{Declaration of Interests}
The authors report no conflict of interest.

%% file: Z_Appendices.tex
\appendix
\input{Z0_Algos}
\input{Z1_Params}

%% file: Z0_Algos.tex
\section{PPO, PPO-CMA and S-PPO-CMA learning algorithms}\label{app:PPOCMA}
Given a partial state $s_t$ and under a policy $\pi$, the advantage value $A^\pi(a_t,s_t)$ compares the value ($R_t$) of a specific action $a_t$ with the expected value of a randomly selected action according to $\pi(\cdot,s_t)$. The latter is simply $V^\pi(s_t)$, the state value computed by the critic neural network $V$, which is an estimator of the expected return $\Expec{\pi}{\sum_\tau \gamma^\tau r_\tau}$ following policy $\pi$. The advantage estimates "how much better" is $a_t$ compared to the "average" action sampled following $\pi$: $A^\pi(a_t,s_t)= R_t-V^\pi(s_t)$. A more stable method of estimating advantage, Generalised Advantage Estimation (GAE) \citep{SchulmanMoritzLevineEtAl2015}, depending on an extra parameter $\lambda_{GAE}$, is used here. PPO and its variants rely on the estimation of the advantage function and use advantage values as weights for the computation of gradients. Concerning PPO, the policy $\pi_\theta(\cdot,s_t)$ follows a random normal distribution whose mean $\mu_\theta(s_t)$ is the output of a neural network $\pi$ ($\theta$ being its weights/biases) and standard deviation $\sigma$ is a predefined hyper-parameter. The surrogate loss computed during policy update is:
\begin{equation}
    \mathcal{L}_{PPO}(t) = \min\left(r_\theta,\text{clip}(r_\theta,1-\varepsilon,1+\varepsilon)\right)A^\pi(a_t,s_t)
\end{equation}
\begin{equation}
    \text{with }r_\theta(t) = \frac{\pi_\theta(a_t,s_t)}{\pi_{\theta_{old}}(a_t,s_t)}\quad \text{and clip}(a,b,c) = \min\left(\max\left(a,b\right),c\right)
\end{equation}
with $\theta_{old}$ being the weights of $\pi$ before update and $\varepsilon$ a clipping hyper-parameter. Algorithm \ref{algo:PPO} describes the steps of PPO, with $\phi$ being the weights and biases of $V$.
\input{algo/PPO}

PPO-CMA \citep{HaemaelaeinenBabadiMaEtAl2018} relies on a similar loss estimation technique, but uses two unclipped surrogate objectives. The standard deviation of the policy $\pi_\theta$ is, this time, also an output of the actor $\pi$. The latter is then trained twice per update phase using two different losses $\mathcal{L}_{PPOCMA}^\sigma$ and $\mathcal{L}_{PPOCMA}^\mu$. PPO being known for instability in policy updates due to negative advantages, PPO-CMA uses a mirroring technique to consider the information brought by negative advantage samples:
\begin{equation}
    \mathcal{L}_{PPOCMA}^\sigma(t) = \delta_{A^\pi>0}A^\pi(a_t,s_t)\pi_\theta(a_t,s_t)
\end{equation}
\begin{equation}
    \mathcal{L}_{PPOCMA}^\mu(t) = \delta_{A^\pi>0}A^\pi(a_t,s_t)\pi_\theta(a_t,s_t) - \delta_{A^\pi<0}A^\pi(a_t,s_t)\pi_\theta(\mu_t-2a_t,s_t)Z(a_t-\mu_t)
\end{equation}
\begin{equation}
    \text{with }\delta_{f(x)}= \left\{\begin{matrix}1 &\text{ if } f(x)\text{ is True}\\
    0 &\text{ otherwise}\end{matrix}\right.
\end{equation}
where $Z$ is a Gaussian kernel damping function, that vanishes when $||a_t-\mu_t||\rightarrow \infty$. To ensure a more stable convergence of $\sigma$, $\mathcal{L}_{PPOCMA}^\sigma$ is estimated on a randomly sampled history buffer $\mathcal{B}$ that contains all the information of the past $H$ epochs of training. Algorithm \ref{algo:PPOCMA} describes the steps of PPO-CMA.
\input{algo/PPOCMA}

The training of S-PPO-CMA (described in section \ref{sec:SPPOCMA}) is similar the standard PPO-CMA. An additional loss $\mathcal{L}_{\text{sparse}}$ is defined to train $\vec{\alpha}$ values. This contains a Tikhonov matrix $\mat{\Gamma}$, that penalises correlations between observations. $\mat{\Gamma}$ is diagonal and for a predefined correlation threshold $\delta_{corr}$:
\begin{equation}
C_i \equiv \{j\neq i\;|\;\text{corr}(s_i,s_j)>\delta_{corr} \}
\end{equation}
\begin{equation}
\Gamma_{ii} = \left\{\begin{matrix}1& \text{ if } C_i \neq \emptyset\text{ and }\exists\;j\in C_i\; \alpha_j > \alpha_i  \\ 0&\text{ otherwise}\end{matrix}\right.
\end{equation}
where $\text{corr}(s_i,s_j)$ is the correlation of $s_i$ and $s_j$ over the current epoch.

The substitute vector $\vec{\bar{s}}$ can be seen as a baseline. There is an advantage in gradient accuracy to choose a slowly updated average of the observation vector as baseline. Let us consider:
\begin{equation}
    \nabla_{\alpha}\mathcal{L}_{\pi_s}(\theta_s,\vec{\alpha}) = \underbrace{\text{sign}\left[\mu_s(\vec{s},\theta_s,\vec{\alpha}) - \mu^*(\vec{s})\right]}_{\delta}\nabla_\alpha\mu_s\left(\vec{s},\theta_s,\vec{\alpha}\right)
\end{equation}
\begin{equation}
    \nabla_{\alpha}\mathcal{L}_{\pi_s}(\theta_s,\vec{\alpha}) = \delta \nabla_{\tilde{\vec{s}}} \pi_{s,\sigma}\left(\theta_s,\tilde{\vec{s}}\right)\odot\nabla_\alpha \tilde{\vec{s}} \quad \text{ with }\tilde{\vec{s}} = (\vec{s}-\bar{\vec{s}})\odot f(\vec{u},\vec{\alpha})+\bar{\vec{s}}
\end{equation}
\begin{equation}
    \nabla_\alpha \tilde{\vec{s}} = (\vec{s}-\bar{\vec{s}})\odot\nabla_\alpha f(\vec{u},\vec{\alpha})
\end{equation}
\begin{equation}
    \text{Thus :}\quad \nabla_{\alpha}\mathcal{L}_{\pi_s}(\theta_s,\vec{\alpha}) = \delta \nabla_{\tilde{\vec{s}}} \pi_{s,\sigma}\left(\theta_s,\tilde{\vec{s}}\right)\odot\nabla_\alpha f(\vec{u},\vec{\alpha})\odot (\vec{s}-\bar{\vec{s}})
\end{equation}
If $\bar{\vec{s}} = 0$, for the values of $\vec{u}$ where $\nabla_\alpha f(\vec{u},\vec{\alpha})\neq0$, the amplitude of the gradient is proportional to the average value of $\vec{s}$. If there is an important disparity in observation averages (\textit{i.e.} $avg(s_i)\gg avg(s_j)$), $\alpha_i$ will be updated much faster than $\alpha_j$ which is not wished. Furthermore, for "small" batches, any correction based on the batch average (using $avg(\vec{s})$ instead of $\bar{\vec{s}}$) introduces a bias due to the lack of convergence of the epoch-averaged estimator $avg$. A slowly updated baseline for $\bar{\vec{s}}$, is thus necessary.

%% file: algo/PPO.tex
\begin{algorithm}
\caption{Proximal Policy Optimisation algorithm}
\label{algo:PPO}
\begin{algorithmic}
\STATE $k \gets 0$
\STATE Initialise $\phi_0$ and $\theta_0$
\WHILE{$\theta_k$ is not converged}
    \STATE Collect a set of trajectories $\mathcal{D}_k=\{\tau_i\}=\{(s,a,r)_i\}$ of length $T$ using policy $\pi_{\theta_k}$
    \FORALL{$t \in [0,T]$}
        \STATE Estimate rewards-to-go $\hat{R}_t = \sum_{t'=t}^T \gamma^{t'-t}r_{t'}$
        \STATE Estimate advantage $\hat{A}_t$ through GAE
    \ENDFOR
    \STATE Estimate policy gradient $\hat{g}=\nabla_\theta \frac{1}{|\mathcal{D}_k|}\sum_{\tau\in \mathcal{D}_k}\sum_{t=0}^{T}\mathcal{L}_{PPO}(a_t,s_t,\hat{A}_t)$
    \STATE Compute policy update $\theta_{k+1}\gets \theta_k + \alpha \hat{g}_k$ (or other gradient technique)
    \STATE Fit value function by regression: $\phi_{k+1}\gets \argmin_\phi \frac{1}{|\mathcal{D}_k|T}\sum_{\tau \in \mathcal{D}_k}\sum_{t=0}^{T}\left(V_\phi(s_t) - \hat{R}_t\right)^2$
    \STATE $k \gets k+1$
\ENDWHILE
\end{algorithmic}
\end{algorithm}

%% file: algo/PPOCMA.tex
\begin{algorithm}
\caption{Proximal Policy Optimisation algorithm with Covariance Matrix Adaptation}
\label{algo:PPOCMA}
\begin{algorithmic}
\STATE $k \gets 0$
\STATE Initialise $\phi_0$ and $\theta_0$
\WHILE{$\theta_k$ is not converged}
    \STATE Collect a set of trajectories $\mathcal{D}_k=\{\tau_i\}=\{(s,a,r)_i\}$ of length $T$ using policy $\pi_{\theta_k}$
    \FORALL{$t \in [0,T]$}
        \STATE Estimate rewards-to-go $\hat{R}_t = \sum_{t'=t}^T \gamma^{t'-t}r_{t'}$
        \STATE Estimate advantage $\hat{A}_t$ through GAE
    \ENDFOR
    \STATE Append $\mathcal{D}_k$, $\hat{R}$ and $\hat{A}$ to the history buffer $\mathcal{B}$
    \STATE Sample $\mathcal{D}'_k$, $\hat{R}'$ and $\hat{A}'$ on history buffer $\mathcal{B}$
    \STATE Estimate $\sigma$  policy gradient $\hat{g}^\sigma=\nabla_\theta \frac{1}{|\mathcal{D}'_k|}\sum_{\tau'\in \mathcal{D}'_k}\sum_{t=0}^{|\mathcal{D}'_k|}\mathcal{L}_{PPOCMA}^\sigma(a'_t,s'_t,\hat{A}'_t)$
    \STATE Compute policy update $\theta_{k+1}\gets \theta_k + \alpha \hat{g}^\sigma_k$ (or other gradient technique)
    \STATE Estimate $\mu$  policy gradient $\hat{g}^\mu=\nabla_\theta \frac{1}{|\mathcal{D}_k|}\sum_{\tau\in \mathcal{D}_k}\sum_{t=0}^{|\mathcal{D}_k|}\mathcal{L}_{PPOCMA}^\mu(a_t,s_t,\hat{A}_t)$
    \STATE Compute policy update $\theta_{k+1}\gets \theta_k + \alpha \hat{g}^\mu_k$ (or other gradient technique)
    \STATE Fit value function by regression: $\phi_{k+1}\gets \argmin_\phi \frac{1}{|\mathcal{D}_k|T}\sum_{\tau \in \mathcal{D}_k}\sum_{t=0}^{T}\left(V_\phi(s_t) - \hat{R}_t\right)^2$
    \STATE $k \gets k+1$
\ENDWHILE
\end{algorithmic}
\end{algorithm}

%% file: Z1_Params.tex
\section{Numerical hyper-parameters}\label{app:params}
Table \ref{tab:params} presents the main numerical parameters of both the simulated case and the learning algorithm, with the notations used in the article (if introduced).
\begin{table}
\centering
\begin{tabular}{cccc}
 Parameter & Symbol & Value & Comment/Reference \\
 \hline
 \multicolumn{4}{c}{\textbf{Flow simulation setup}} \\
Spatial scheme & - & AUSM+ & \citet{Edwards1998} \\
Mesh nodes (aziumtally) & - & 360 & - \\
Mesh nodes (radially) & - & 70 & - \\
Temporal scheme & - & BDF2 & \citet{Curtiss1952} \\
Numerical time step & $dt$ & $5\times10^{-3}$ & -\\
Action ramp length & - & $20$ it. & - \\
Maximum action amplitude & - & $2$ & - \\
Control step length & $\Delta t$& $50$ it. $=0.25$ & - \\
\hline
\multicolumn{4}{c}{\textbf{(S)-PPO-CMA hyper-parameters}}\\
Training epochs & - & $200$ & - \\
Steps per epoch & - & $480$ & -  \\
Actor architecture & $\pi$ & $(512\times512)$ & 2 fully connected layers\\
Critic architecture & $V$ & $(512\times512)$ & 2 fully connected layers\\
Return discount factor & $\gamma$& $0.99$ & - \\
GAE control parameter & $\lambda_{GAE}$ & $0.97$ & Standard value\\
Optimiser & - & ADAM & \citet{Kingma2014}\\
History buffer depth & $H$ & $3$ epochs & - \\
Bernoulli choice parameter on action & $p$ & $0.2$ & - \\
$L_0$ regularisation parameter & $\lambda$ & $[0.1;10]$ & - \\
Correlation threshold & $\delta_{corr}$ & $0.99$ & - \\
\end{tabular}
\caption{\label{tab:params} Additional numerical parameters}
\end{table}

%% file: main.bbl
\begin{thebibliography}{76}
\expandafter\ifx\csname natexlab\endcsname\relax\def\natexlab#1{#1}\fi
\def\au#1{#1} \def\ed#1{#1} \def\yr#1{#1}\def\at#1{#1}\def\jt#1{\textit{#1}}
  \def\bt#1{#1}\def\bvol#1{\textbf{#1}} \def\vol#1{#1} \def\pg#1{#1}
  \def\publ#1{#1}\def\arxiv#1{#1}\def\org#1{#1}\def\st#1{\textit{#1}}

\bibitem[Abadi {\em et~al.\/}(2016)Abadi, Barham, Chen, Chen, Davis, Dean,
  Devin, Ghemawat, Irving \& Isard]{Abadi2016}
{\sc \au{Abadi, Martín}, \au{Barham, Paul}, \au{Chen, Jianmin}, \au{Chen,
  Zhifeng}, \au{Davis, Andy}, \au{Dean, Jeffrey}, \au{Devin, Matthieu},
  \au{Ghemawat, Sanjay}, \au{Irving, Geoffrey} \& \au{Isard, Michael}}
  \yr{2016}  \at{Tensorflow: A system for large-scale machine learning}.
  \jt{12th {USENIX} Symposium on Operating Systems Design and Implementation
  ({OSDI} 16)}  \pg{pp. 265--283}.

\bibitem[Arakeri \& Shukla(2013)]{Arakeri2013}
{\sc \au{Arakeri, Jaywant~H.} \& \au{Shukla, Ratnesh~K.}} \yr{2013}  \at{A
  unified view of energetic efficiency in active drag reduction, thrust
  generation and self-propulsion through a loss coefficient with some
  applications}.  \jt{Journal of Fluids and Structures}  \bvol{41},
  \pg{22--32}.

\bibitem[Baker {\em et~al.\/}(2019)Baker, Kanitscheider, Markov, Wu, Powell,
  McGrew \& Mordatch]{Baker2019}
{\sc \au{Baker, Bowen}, \au{Kanitscheider, Ingmar}, \au{Markov, Todor}, \au{Wu,
  Yi}, \au{Powell, Glenn}, \au{McGrew, Bob} \& \au{Mordatch, Igor}} \yr{2019}
  \at{Emergent tool use from multi-agent autocurricula}.  \jt{arXiv preprint
  arXiv: 1909.07528} .

\bibitem[Barkley(2006)]{Barkley2006}
{\sc \au{Barkley, D.}} \yr{2006}  \at{Linear analysis of the cylinder wake mean
  flow}.  \jt{EPL (Europhysics Letters)}  \bvol{75}~(5),  \pg{750}.

\bibitem[Beneddine(2017)]{Beneddine2017}
{\sc \au{Beneddine, Samir}} \yr{2017}  \at{Characterization of unsteady flow
  behavior by linear stability analysis}. PhD thesis.

\bibitem[Benoit {\em et~al.\/}(2015)Benoit, Péron \& Landier]{Benoit2015}
{\sc \au{Benoit, Christophe}, \au{Péron, Stéphanie} \& \au{Landier, Sâm}}
  \yr{2015}  \at{{Cassiopee: a CFD pre-and post-processing tool}}.
  \jt{Aerospace Science and Technology}  \bvol{45},  \pg{272--283}.

\bibitem[Bergmann \& Cordier(2008)]{Bergmann2008}
{\sc \au{Bergmann, Michel} \& \au{Cordier, Laurent}} \yr{2008}  \at{Optimal
  control of the cylinder wake in the laminar regime by trust-region methods
  and pod reduced-order models}.  \jt{Journal of Computational Physics}
  \bvol{227}~(16),  \pg{7813--7840}.

\bibitem[Bergmann {\em et~al.\/}(2005)Bergmann, Cordier \&
  Brancher]{Bergmann2005}
{\sc \au{Bergmann, Michel}, \au{Cordier, Laurent} \& \au{Brancher, J.-P.}}
  \yr{2005}  \at{Control of the cylinder wake in the laminar regime by
  trust-region methods and pod reduced order models}.  \jt{Proceedings of the
  44th IEEE Conference on Decision and Control}  \pg{pp. 524--529}.

\bibitem[Bergmann {\em et~al.\/}(2006)Bergmann, Cordier \&
  Brancher]{Bergmann2006}
{\sc \au{Bergmann, Michel}, \au{Cordier, Laurent} \& \au{Brancher,
  Jean-Pierre}} \yr{2006}  \at{On the generation of a reverse von kármán
  street for the controlled cylinder wake in the laminar regime}.  \jt{Physics
  of Fluids}  \bvol{18}~(2),  \pg{028101}.

\bibitem[Braza {\em et~al.\/}(1986)Braza, Chassaing \& Minh]{Braza1986}
{\sc \au{Braza, M.}, \au{Chassaing, P. H. H.~M.} \& \au{Minh, H.~Ha}} \yr{1986}
   \at{Numerical study and physical analysis of the pressure and velocity
  fields in the near wake of a circular cylinder}.  \jt{Journal of Fluid
  Mechanics}  \bvol{165},  \pg{79--130}.

\bibitem[Bright {\em et~al.\/}(2013)Bright, Lin \& Kutz]{Bright2013}
{\sc \au{Bright, Ido}, \au{Lin, Guang} \& \au{Kutz, J.~Nathan}} \yr{2013}
  \at{Compressive sensing based machine learning strategy for characterizing
  the flow around a cylinder with limited pressure measurements}.  \jt{Physics
  of Fluids}  \bvol{25}~(12),  \pg{127102}.

\bibitem[Brockman {\em et~al.\/}(2016)Brockman, Cheung, Pettersson, Schneider,
  Schulman, Tang \& Zaremba]{Brockman2016}
{\sc \au{Brockman, Greg}, \au{Cheung, Vicki}, \au{Pettersson, Ludwig},
  \au{Schneider, Jonas}, \au{Schulman, John}, \au{Tang, Jie} \& \au{Zaremba,
  Wojciech}} \yr{2016}  \at{{OpenAI Gym}}.  \jt{arXiv preprint
  arXiv:1606.01540} .

\bibitem[Brunton \& Noack(2015)]{BruntonNoack2015}
{\sc \au{Brunton, Steven~L} \& \au{Noack, Bernd~R}} \yr{2015}  \at{Closed-loop
  turbulence control: progress and challenges}.  \jt{Applied Mechanics Reviews}
   \bvol{67}~(5),  \pg{050801--}.

\bibitem[Brunton {\em et~al.\/}(2020)Brunton, Noack \&
  Koumoutsakos]{Brunton2020}
{\sc \au{Brunton, Steven~L.}, \au{Noack, Bernd~R.} \& \au{Koumoutsakos,
  Petros}} \yr{2020}  \at{Machine learning for fluid mechanics}.  \jt{Annual
  Review of Fluid Mechanics}  \bvol{52},  \pg{477--508}.

\bibitem[Chen \& Aubry(2005)]{Chen2005}
{\sc \au{Chen, Zhihua} \& \au{Aubry, Nadine}} \yr{2005}  \at{Active control of
  cylinder wake}.  \jt{Communications in nonlinear science and numerical
  simulation}  \bvol{10}~(2),  \pg{205--216}.

\bibitem[Cho {\em et~al.\/}(2014)Cho, Van~Merriënboer, Gulcehre, Bahdanau,
  Bougares, Schwenk \& Bengio]{Cho2014}
{\sc \au{Cho, Kyunghyun}, \au{Van~Merriënboer, Bart}, \au{Gulcehre, Caglar},
  \au{Bahdanau, Dzmitry}, \au{Bougares, Fethi}, \au{Schwenk, Holger} \&
  \au{Bengio, Yoshua}} \yr{2014}  \at{Learning phrase representations using rnn
  encoder-decoder for statistical machine translation}.  \jt{arXiv preprint
  arXiv: 1406.1078} .

\bibitem[Cohen {\em et~al.\/}(2006)Cohen, Siegel \& McLaughlin]{Cohen2006}
{\sc \au{Cohen, Kelly}, \au{Siegel, Stefan} \& \au{McLaughlin, Thomas}}
  \yr{2006}  \at{A heuristic approach to effective sensor placement for
  modeling of a cylinder wake}.  \jt{Computers \& fluids}  \bvol{35}~(1),
  \pg{103--120}.

\bibitem[Cohen {\em et~al.\/}(2012)Cohen, Siegel, Seidel, Aradag \&
  McLaughlin]{Cohen2012}
{\sc \au{Cohen, Kelly}, \au{Siegel, Stefan}, \au{Seidel, Jürgen}, \au{Aradag,
  Selin} \& \au{McLaughlin, Thomas}} \yr{2012}  \at{Nonlinear estimation of
  transient flow field low dimensional states using artificial neural nets}.
  \jt{Expert Systems with Applications}  \bvol{39}~(1),  \pg{1264--1272}.

\bibitem[Curtiss \& Hirschfelder(1952)]{Curtiss1952}
{\sc \au{Curtiss, Charles~Francis} \& \au{Hirschfelder, Joseph~O.}} \yr{1952}
  \at{Integration of stiff equations}.  \jt{Proceedings of the National Academy
  of Sciences of the United States of America}  \bvol{38}~(3),  \pg{235}.

\bibitem[Dandois {\em et~al.\/}(2013)Dandois, Garnier \&
  Pamart]{DandoisGarnierPamart2013}
{\sc \au{Dandois, J.}, \au{Garnier, E.} \& \au{Pamart, P.-Y.}} \yr{2013}
  \at{{NARX modelling of unsteady separation control}}.  \jt{Experiments in
  fluids}  \bvol{54}~(2),  \pg{1445}.

\bibitem[Dandois {\em et~al.\/}(2018)Dandois, Mary \& Brion]{Dandois2018}
{\sc \au{Dandois, Julien}, \au{Mary, Ivan} \& \au{Brion, Vincent}} \yr{2018}
  \at{Large-eddy simulation of laminar transonic buffet}.  \jt{Journal of Fluid
  Mechanics}  \bvol{850},  \pg{156--178}.

\bibitem[DeVries \& Paley(2013)]{DeVries2013}
{\sc \au{DeVries, Levi} \& \au{Paley, Derek~A.}} \yr{2013}
  \at{Observability-based optimization for flow sensing and control of an
  underwater vehicle in a uniform flowfield}.  \jt{2013 American Control
  Conference}  \pg{pp. 1386--1391}.

\bibitem[Edwards \& Liou(1998)]{Edwards1998}
{\sc \au{Edwards, Jack~R.} \& \au{Liou, Meng-Sing}} \yr{1998}
  \at{Low-diffusion flux-splitting methods for flows at all speeds}.  \jt{AIAA
  Journal}  \bvol{36}~(9),  \pg{1610--1617}.

\bibitem[Foures {\em et~al.\/}(2014)Foures, Dovetta, Sipp \&
  Schmid]{Foures2014}
{\sc \au{Foures, Dimitry P.~G.}, \au{Dovetta, Nicolas}, \au{Sipp, Denis} \&
  \au{Schmid, Peter~J.}} \yr{2014}  \at{{A data-assimilation method for
  Reynolds-averaged Navier-Stokes-driven mean flow reconstruction}}.
  \jt{Journal of Fluid Mechanics}  \bvol{759},  \pg{404--431}.

\bibitem[Fujisawa {\em et~al.\/}(2001)Fujisawa, Kawaji \&
  Ikemoto]{Fujisawa2001}
{\sc \au{Fujisawa, N.}, \au{Kawaji, Y.} \& \au{Ikemoto, K.}} \yr{2001}
  \at{Feedback control of vortex shedding from a circular cylinder by
  rotational oscillations}.  \jt{Journal of Fluids and Structures}
  \bvol{15}~(1),  \pg{23--37}.

\bibitem[Gerhard {\em et~al.\/}(2003)Gerhard, Pastoor, King, Noack, Dillmann,
  Morzynski \& Tadmor]{Gerhard2003}
{\sc \au{Gerhard, Johannes}, \au{Pastoor, Mark}, \au{King, Rudibert},
  \au{Noack, Bernd}, \au{Dillmann, Andreas}, \au{Morzynski, Marek} \&
  \au{Tadmor, Gilead}} \yr{2003}  \at{Model-based control of vortex shedding
  using low-dimensional galerkin models}.  \jt{33rd AIAA Fluid Dynamics
  Conference and Exhibit}  \pg{p. 4262}.

\bibitem[Gutmark \& Grinstein(1999)]{Gutmark1999}
{\sc \au{Gutmark, E.~J.} \& \au{Grinstein, F.~F.}} \yr{1999}  \at{Flow control
  with noncircular jets}.  \jt{Annual review of fluid mechanics}
  \bvol{31}~(1),  \pg{239--272}.

\bibitem[Hansen(2016)]{Hansen2016}
{\sc \au{Hansen, Nikolaus}} \yr{2016}  \at{The cma evolution strategy: A
  tutorial}.  \jt{arXiv preprint arXiv: 1604.00772} .

\bibitem[Hansen {\em et~al.\/}(2003)Hansen, Müller \&
  Koumoutsakos]{Hansen2003}
{\sc \au{Hansen, Nikolaus}, \au{Müller, Sibylle~D.} \& \au{Koumoutsakos,
  Petros}} \yr{2003}  \at{{Reducing the time complexity of the derandomized
  evolution strategy with covariance matrix adaptation (CMA-ES)}}.
  \jt{Evolutionary computation}  \bvol{11}~(1),  \pg{1--18}.

\bibitem[He {\em et~al.\/}(2000)He, Glowinski, Metcalfe, Nordlander \&
  Periaux]{He2000}
{\sc \au{He, J.-W.}, \au{Glowinski, R.}, \au{Metcalfe, R.}, \au{Nordlander, A.}
  \& \au{Periaux, J.}} \yr{2000}  \at{Active control and drag optimization for
  flow past a circular cylinder: I. oscillatory cylinder rotation}.
  \jt{Journal of Computational Physics}  \bvol{163}~(1),  \pg{83--117}.

\bibitem[He {\em et~al.\/}(2016)He, Zhang, Ren \& Sun]{He2016}
{\sc \au{He, Kaiming}, \au{Zhang, Xiangyu}, \au{Ren, Shaoqing} \& \au{Sun,
  Jian}} \yr{2016}  \at{Deep residual learning for image recognition}.
  \jt{Proceedings of the IEEE conference on computer vision and pattern
  recognition}  \pg{pp. 770--778}.

\bibitem[Henderson(1997)]{Henderson1997}
{\sc \au{Henderson, Ronald~D.}} \yr{1997}  \at{Nonlinear dynamics and pattern
  formation in turbulent wake transition}.  \jt{Journal of Fluid Mechanics}
  \bvol{352},  \pg{65--112}.

\bibitem[Huh {\em et~al.\/}(2016)Huh, Agrawal \& Efros]{Huh2016}
{\sc \au{Huh, Minyoung}, \au{Agrawal, Pulkit} \& \au{Efros, Alexei~A.}}
  \yr{2016}  \at{What makes imagenet good for transfer learning?}  \jt{arXiv
  preprint arXiv: 1608.08614} .

\bibitem[Hämäläinen {\em et~al.\/}(2018)Hämäläinen, Babadi, Ma \&
  Lehtinen]{HaemaelaeinenBabadiMaEtAl2018}
{\sc \au{Hämäläinen, Perttu}, \au{Babadi, Amin}, \au{Ma, Xiaoxiao} \&
  \au{Lehtinen, Jaakko}} \yr{2018}  \at{{PPO-CMA: proximal policy optimization
  with covariance matrix adaptation}}.  \jt{arXiv preprint arXiv: 1810.02541} .

\bibitem[Jin {\em et~al.\/}(2019)Jin, Illingworth \& Sandberg]{Jin2019}
{\sc \au{Jin, Bo}, \au{Illingworth, Simon~J.} \& \au{Sandberg, Richard~D.}}
  \yr{2019}  \at{Feedback control of vortex shedding using a resolvent-based
  modelling approach}.  \jt{arXiv preprint arXiv:1909.04865} .

\bibitem[Kaiser {\em et~al.\/}(2019)Kaiser, Babaeizadeh, Milos, Osinski,
  Campbell, Czechowski, Erhan, Finn, Kozakowski \& Levine]{Kaiser2019}
{\sc \au{Kaiser, Lukasz}, \au{Babaeizadeh, Mohammad}, \au{Milos, Piotr},
  \au{Osinski, Blazej}, \au{Campbell, Roy~H.}, \au{Czechowski, Konrad},
  \au{Erhan, Dumitru}, \au{Finn, Chelsea}, \au{Kozakowski, Piotr} \&
  \au{Levine, Sergey}} \yr{2019}  \at{{Model-based reinforcement learning for
  Atari}}.  \jt{arXiv preprint arXiv:1903.00374} .

\bibitem[Kim {\em et~al.\/}(2006)Kim, Kerr, Beskok \& Jayasuriya]{Kim2006}
{\sc \au{Kim, Kihwan}, \au{Kerr, Murray}, \au{Beskok, Ali} \& \au{Jayasuriya,
  Suhada}} \yr{2006}  \at{Frequency-domain based feedback control of flow
  separation using synthetic jets}.  \jt{2006 American Control Conference}
  \pg{pp. 6--pp}.

\bibitem[Kingma \& Ba(2014)]{Kingma2014}
{\sc \au{Kingma, Diederik~P.} \& \au{Ba, Jimmy}} \yr{2014}  \at{Adam: A method
  for stochastic optimization}.  \jt{arXiv preprint arXiv:1412.6980} .

\bibitem[Leclerc {\em et~al.\/}(2006)Leclerc, Sagaut \& Mohammadi]{Leclerc2006}
{\sc \au{Leclerc, Eric}, \au{Sagaut, Pierre} \& \au{Mohammadi, Bijan}}
  \yr{2006}  \at{On the use of incomplete sensitivities for feedback control of
  laminar vortex shedding}.  \jt{Computers \& fluids}  \bvol{35}~(10),
  \pg{1432--1443}.

\bibitem[Leclercq {\em et~al.\/}(2019)Leclercq, Demourant, Poussot-Vassal \&
  Sipp]{LeclercqDemourantPoussot-VassalEtAl2019}
{\sc \au{Leclercq, Colin}, \au{Demourant, Fabrice}, \au{Poussot-Vassal,
  Charles} \& \au{Sipp, Denis}} \yr{2019}  \at{Linear iterative method for
  closed-loop control of quasiperiodic flows}.  \jt{Journal of Fluid Mechanics}
   \bvol{868},  \pg{26--65}.

\bibitem[Louizos {\em et~al.\/}(2017)Louizos, Welling \& Kingma]{Louizos2017}
{\sc \au{Louizos, Christos}, \au{Welling, Max} \& \au{Kingma, Diederik~P.}}
  \yr{2017}  \at{Learning sparse neural networks through $ l_0 $
  regularization}.  \jt{arXiv preprint arXiv:1712.01312} .

\bibitem[Manohar {\em et~al.\/}(2018)Manohar, Kutz \& Brunton]{Manohar2018}
{\sc \au{Manohar, Krithika}, \au{Kutz, J.~Nathan} \& \au{Brunton, Steven~L.}}
  \yr{2018}  \at{Optimal sensor and actuator placement using balanced model
  reduction}.  \jt{arXiv preprint arXiv:1812.01574} .

\bibitem[Marquet {\em et~al.\/}(2008)Marquet, Sipp \& Jacquin]{Marquet2008}
{\sc \au{Marquet, Olivier}, \au{Sipp, Denis} \& \au{Jacquin, Laurent}}
  \yr{2008}  \at{Sensitivity analysis and passive control of cylinder flow}.
  \jt{Journal of Fluid Mechanics}  \bvol{615},  \pg{221--252}.

\bibitem[Min \& Choi(1999)]{Min1999}
{\sc \au{Min, Chulhong} \& \au{Choi, Haecheon}} \yr{1999}  \at{Suboptimal
  feedback control of vortex shedding at low reynolds numbers}.  \jt{Journal of
  Fluid Mechanics}  \bvol{401},  \pg{123--156}.

\bibitem[Mnih {\em et~al.\/}(2015)Mnih, Kavukcuoglu, Silver, Rusu, Veness,
  Bellemare, Graves, Riedmiller, Fidjeland \&
  Ostrovski]{MnihKavukcuogluSilverEtAl2015}
{\sc \au{Mnih, Volodymyr}, \au{Kavukcuoglu, Koray}, \au{Silver, David},
  \au{Rusu, Andrei~A.}, \au{Veness, Joel}, \au{Bellemare, Marc~G.}, \au{Graves,
  Alex}, \au{Riedmiller, Martin}, \au{Fidjeland, Andreas~K.} \& \au{Ostrovski,
  Georg}} \yr{2015}  \at{Human-level control through deep reinforcement
  learning}.  \jt{Nature}  \bvol{518}~(7540),  \pg{529}.

\bibitem[Mons {\em et~al.\/}(2016)Mons, Chassaing, Gomez \& Sagaut]{Mons2016}
{\sc \au{Mons, Vincent}, \au{Chassaing, J.-C.}, \au{Gomez, Thomas} \&
  \au{Sagaut, Pierre}} \yr{2016}  \at{Reconstruction of unsteady viscous flows
  using data assimilation schemes}.  \jt{Journal of Computational Physics}
  \bvol{316},  \pg{255--280}.

\bibitem[Mons {\em et~al.\/}(2017)Mons, Chassaing \& Sagaut]{Mons2017}
{\sc \au{Mons, Vincent}, \au{Chassaing, Jean-Camille} \& \au{Sagaut, Pierre}}
  \yr{2017}  \at{Optimal sensor placement for variational data assimilation of
  unsteady flows past a rotationally oscillating cylinder}.  \jt{Journal of
  Fluid Mechanics}  \bvol{823},  \pg{230--277}.

\bibitem[Muddada \& Patnaik(2010)]{Muddada2010}
{\sc \au{Muddada, Sridhar} \& \au{Patnaik, B. S.~V.}} \yr{2010}  \at{An active
  flow control strategy for the suppression of vortex structures behind a
  circular cylinder}.  \jt{European Journal of Mechanics-B/Fluids}
  \bvol{29}~(2),  \pg{93--104}.

\bibitem[Nair {\em et~al.\/}(2020)Nair, Taira, Brunton \& Brunton]{Nair2020}
{\sc \au{Nair, Aditya~G.}, \au{Taira, Kunihiko}, \au{Brunton, Bingni~W.} \&
  \au{Brunton, Steven~L.}} \yr{2020}  \at{Phase-based control of periodic fluid
  flows}.  \jt{arXiv preprint arXiv:2004.10561} .

\bibitem[Nishioka \& Sato(1978)]{Nishioka1978}
{\sc \au{Nishioka, Michio} \& \au{Sato, Hiroshi}} \yr{1978}  \at{Mechanism of
  determination of the shedding frequency of vortices behind a cylinder at low
  reynolds numbers}.  \jt{Journal of Fluid Mechanics}  \bvol{89}~(1),
  \pg{49--60}.

\bibitem[Nørgård {\em et~al.\/}(2000)Nørgård, Ravn, Poulsen \&
  Hansen]{Noergaard2000}
{\sc \au{Nørgård, Peter~Magnus}, \au{Ravn, Ole}, \au{Poulsen,
  Niels~Kjølstad} \& \au{Hansen, Lars~Kai}} \yr{2000} {\em Neural networks for
  modelling and control of dynamic systems-A practitioner's handbook\/}.
  \publ{Springer-London}.

\bibitem[Oehler \& Illingworth(2018)]{Oehler2018}
{\sc \au{Oehler, Stephan~F.} \& \au{Illingworth, Simon~J.}} \yr{2018}
  \at{Sensor and actuator placement trade-offs for a linear model of spatially
  developing flows}.  \jt{Journal of Fluid Mechanics}  \bvol{854},
  \pg{34--55}.

\bibitem[Protas \& Styczek(2002)]{Protas2002a}
{\sc \au{Protas, B.} \& \au{Styczek, A.}} \yr{2002}  \at{Optimal rotary control
  of the cylinder wake in the laminar regime}.  \jt{Physics of Fluids}
  \bvol{14}~(7),  \pg{2073--2087}.

\bibitem[Protas \& Wesfreid(2002)]{Protas2002}
{\sc \au{Protas, B.} \& \au{Wesfreid, J.~E.}} \yr{2002}  \at{Drag force in the
  open-loop control of the cylinder wake in the laminar regime}.  \jt{Physics
  of Fluids}  \bvol{14}~(2),  \pg{810--826}.

\bibitem[Rabault {\em et~al.\/}(2019)Rabault, Kuchta, Jensen, Réglade \&
  Cerardi]{RabaultKuchtaJensenEtAl2019}
{\sc \au{Rabault, Jean}, \au{Kuchta, Miroslav}, \au{Jensen, Atle},
  \au{Réglade, Ulysse} \& \au{Cerardi, Nicolas}} \yr{2019}  \at{Artificial
  neural networks trained through deep reinforcement learning discover control
  strategies for active flow control}.  \jt{Journal of Fluid Mechanics}
  \bvol{865},  \pg{281--302}.

\bibitem[Rabault \& Kuhnle(2019)]{RabaultKuhnle2019}
{\sc \au{Rabault, Jean} \& \au{Kuhnle, Alexander}} \yr{2019}  \at{Accelerating
  deep reinforcement learning strategies of flow control through a
  multi-environment approach}.  \jt{Physics of Fluids}  \bvol{31}~(9),
  \pg{094105}.

\bibitem[Rabault {\em et~al.\/}(2020)Rabault, Ren, Zhang, Tang \&
  Xu]{RabaultRenZhangEtAl2020}
{\sc \au{Rabault, Jean}, \au{Ren, Feng}, \au{Zhang, Wei}, \au{Tang, Hui} \&
  \au{Xu, Hui}} \yr{2020}  \at{Deep reinforcement learning in fluid mechanics:
  a promising method for both active flow control and shape optimization}.
  \jt{arXiv preprint arXiv:2001.02464} .

\bibitem[Rashidi {\em et~al.\/}(2016)Rashidi, Hayatdavoodi \&
  Esfahani]{Rashidi2016}
{\sc \au{Rashidi, Saman}, \au{Hayatdavoodi, Masoud} \& \au{Esfahani,
  Javad~Abolfazli}} \yr{2016}  \at{Vortex shedding suppression and wake
  control: A review}.  \jt{Ocean Engineering}  \bvol{126},  \pg{57--80}.

\bibitem[Schulman {\em et~al.\/}(2015{\natexlab{{\em a\/}}})Schulman, Levine,
  Abbeel, Jordan \& Moritz]{SchulmanLevineAbbeelEtAl2015}
{\sc \au{Schulman, John}, \au{Levine, Sergey}, \au{Abbeel, Pieter}, \au{Jordan,
  Michael} \& \au{Moritz, Philipp}} \yr{2015{\natexlab{{\em a\/}}}}  \at{Trust
  region policy optimization}.  \jt{International conference on machine
  learning}  \pg{pp. 1889--1897}.

\bibitem[Schulman {\em et~al.\/}(2015{\natexlab{{\em b\/}}})Schulman, Moritz,
  Levine, Jordan \& Abbeel]{SchulmanMoritzLevineEtAl2015}
{\sc \au{Schulman, John}, \au{Moritz, Philipp}, \au{Levine, Sergey},
  \au{Jordan, Michael} \& \au{Abbeel, Pieter}} \yr{2015{\natexlab{{\em b\/}}}}
  \at{High-dimensional continuous control using generalized advantage
  estimation}.  \jt{arXiv preprint arXiv:1506.02438} .

\bibitem[Schulman {\em et~al.\/}(2017)Schulman, Wolski, Dhariwal, Radford \&
  Klimov]{SchulmanWolskiDhariwalEtAl2017}
{\sc \au{Schulman, John}, \au{Wolski, Filip}, \au{Dhariwal, Prafulla},
  \au{Radford, Alec} \& \au{Klimov, Oleg}} \yr{2017}  \at{Proximal policy
  optimization algorithms}.  \jt{arXiv preprint arXiv:1707.06347} .

\bibitem[Seidel {\em et~al.\/}(2009)Seidel, Siegel, Fagley, Cohen \&
  McLaughlin]{Seidel2009}
{\sc \au{Seidel, Jürgen}, \au{Siegel, Stefan}, \au{Fagley, C.}, \au{Cohen, K.}
  \& \au{McLaughlin, T.}} \yr{2009}  \at{Feedback control of a circular
  cylinder wake}.  \jt{Proceedings of the Institution of Mechanical Engineers,
  Part G: Journal of Aerospace Engineering}  \bvol{223}~(4),  \pg{379--392}.

\bibitem[Selby {\em et~al.\/}(1992)Selby, Lin \& Howard]{Selby1992}
{\sc \au{Selby, G.~V.}, \au{Lin, J.~C.} \& \au{Howard, F.~G.}} \yr{1992}
  \at{Control of low-speed turbulent separated flow using jet vortex
  generators}.  \jt{Experiments in Fluids}  \bvol{12}~(6),  \pg{394--400}.

\bibitem[Siegel {\em et~al.\/}(2003)Siegel, Cohen \& McLaughlin]{Siegel2003}
{\sc \au{Siegel, Stefan}, \au{Cohen, Kelly} \& \au{McLaughlin, Tom}} \yr{2003}
  \at{Feedback control of a circular cylinder wake in experiment and
  simulation}.  \jt{33rd AIAA Fluid Dynamics Conference and Exhibit}  \pg{p.
  3569}.

\bibitem[Singh \& Hahn(2005)]{Singh2005}
{\sc \au{Singh, Abhay~K.} \& \au{Hahn, Juergen}} \yr{2005}  \at{Determining
  optimal sensor locations for state and parameter estimation for stable
  nonlinear systems}.  \jt{Industrial \& engineering chemistry research}
  \bvol{44}~(15),  \pg{5645--5659}.

\bibitem[Singha \& Sinhamahapatra(2011)]{Singha2011}
{\sc \au{Singha, Sintu} \& \au{Sinhamahapatra, K.~P.}} \yr{2011}  \at{Control
  of vortex shedding from a circular cylinder using imposed transverse magnetic
  field}.  \jt{International Journal of Numerical Methods for Heat \& Fluid
  Flow}  \bvol{21}~(1),  \pg{32--45}.

\bibitem[Sipp(2012)]{Sipp2012}
{\sc \au{Sipp, Denis}} \yr{2012}  \at{Open-loop control of cavity oscillations
  with harmonic forcings}.  \jt{Journal of Fluid Mechanics}  \bvol{708},
  \pg{439--468}.

\bibitem[Sipp {\em et~al.\/}(2010)Sipp, Marquet, Meliga \&
  Barbagallo]{SippMarquetMeligaEtAl2010}
{\sc \au{Sipp, Denis}, \au{Marquet, Olivier}, \au{Meliga, Philippe} \&
  \au{Barbagallo, Alexandre}} \yr{2010}  \at{Dynamics and control of global
  instabilities in open-flows: a linearized approach}.  \jt{Applied Mechanics
  Reviews}  \bvol{63}~(3).

\bibitem[Sipp \& Schmid(2016)]{Sipp2016}
{\sc \au{Sipp, Denis} \& \au{Schmid, Peter~J.}} \yr{2016}  \at{Linear
  closed-loop control of fluid instabilities and noise-induced perturbations: A
  review of approaches and tools}.  \jt{Applied Mechanics Reviews}
  \bvol{68}~(2).

\bibitem[Sohankar {\em et~al.\/}(2015)Sohankar, Khodadadi \&
  Rangraz]{Sohankar2015}
{\sc \au{Sohankar, A.}, \au{Khodadadi, M.} \& \au{Rangraz, E.}} \yr{2015}
  \at{{Control of fluid flow and heat transfer around a square cylinder by
  uniform suction and blowing at low Reynolds numbers}}.  \jt{Computers \&
  Fluids}  \bvol{109},  \pg{155--167}.

\bibitem[Sutskever {\em et~al.\/}(2014)Sutskever, Vinyals \& Le]{Sutskever2014}
{\sc \au{Sutskever, Ilya}, \au{Vinyals, Oriol} \& \au{Le, Quoc~V.}} \yr{2014}
  \at{Sequence to sequence learning with neural networks}.  \jt{Advances in
  neural information processing systems}  \pg{pp. 3104--3112}.

\bibitem[Tang {\em et~al.\/}(2020)Tang, Rabault, Kuhnle, Wang \&
  Wang]{Tang2020}
{\sc \au{Tang, Hongwei}, \au{Rabault, Jean}, \au{Kuhnle, Alexander}, \au{Wang,
  Yan} \& \au{Wang, Tongguang}} \yr{2020}  \at{Robust active flow control over
  a range of reynolds numbers using an artificial neural network trained
  through deep reinforcement learning}.  \jt{Physics of Fluids}  \bvol{32}~(5),
   \pg{053605}.

\bibitem[Verma {\em et~al.\/}(2020)Verma, Papadimitriou, Lüthen, Arampatzis \&
  Koumoutsakos]{Verma2020}
{\sc \au{Verma, Siddhartha}, \au{Papadimitriou, Costas}, \au{Lüthen, Nora},
  \au{Arampatzis, Georgios} \& \au{Koumoutsakos, Petros}} \yr{2020}
  \at{Optimal sensor placement for artificial swimmers}.  \jt{Journal of Fluid
  Mechanics}  \bvol{884}.

\bibitem[Williams(1992)]{Williams1992}
{\sc \au{Williams, Ronald~J.}} \yr{1992}  \at{Simple statistical
  gradient-following algorithms for connectionist reinforcement learning}.
  \jt{Machine learning}  \bvol{8}~(3-4),  \pg{229--256}.

\bibitem[Williamson(1996)]{Williamson1996}
{\sc \au{Williamson, Charles H.~K.}} \yr{1996}  \at{Vortex dynamics in the
  cylinder wake}.  \jt{Annual review of fluid mechanics}  \bvol{28}~(1),
  \pg{477--539}.

\bibitem[Zielinska {\em et~al.\/}(1997)Zielinska, Goujon-Durand, Dusek \&
  Wesfreid]{Zielinska1997}
{\sc \au{Zielinska, B. J.~A.}, \au{Goujon-Durand, S.}, \au{Dusek, J.} \&
  \au{Wesfreid, J.~E.}} \yr{1997}  \at{Strongly nonlinear effect in unstable
  wakes}.  \jt{Physical review letters}  \bvol{79}~(20),  \pg{3893}.

\end{thebibliography}
